\documentclass[preprint,superscriptaddress,amsmath,amssymb,prd,aps,showpacs,floatfix,nofootinbib]{revtex4-1}
\usepackage{graphicx}
\usepackage{dcolumn}
\usepackage{bm}
\usepackage{mathrsfs}
\usepackage{hyperref}
\usepackage{amsmath}
\usepackage{amssymb}
\usepackage{color}
\usepackage{float}
\usepackage{dcolumn}
\usepackage{multirow}
\usepackage{changepage}

\begin{document}
\title{Molecular $\Xi_{bc}$ states from meson-baryon interaction}
\date{\today}
\author{Qi-Xin~Yu}
\email{yuqx@mail.bnu.edu.cn}
\affiliation{Department of Physics, Guangxi Normal University, Guilin 541004, China}
\affiliation{Institute for Experimental Physics, Department of Physics, University of Hamburg, Luruper Chaussee 149, D-22761 Hamburg, Germany}

\author{J.~M.~Dias}
\email{isengardjor@gmail.com}
\affiliation{Department of Physics, Guangxi Normal University, Guilin 541004, China}
\affiliation{Instituto de F{\'i}sica, Universidade de S{\~a}o Paulo, Rua do Mat{\~a}o 1371. Butant{\~a}, CEP 05508-090, S{\~a}o Paulo, S{\~a}o Paulo, Brazil}

\author{Wei-Hong~Liang}
\email{liangwh@gxnu.edu.cn}
\affiliation{Department of Physics, Guangxi Normal University, Guilin 541004, China}
\affiliation{Guangxi Key Laboratory of Nuclear Physics and Technology, Guangxi Normal University, Guilin 541004, China}

\author{E.~Oset}
\email{oset@ific.uv.es}
\affiliation{Department of Physics, Guangxi Normal University, Guilin 541004, China}
\affiliation{Departamento de
F\'{\i}sica Te\'orica and IFIC, Centro Mixto Universidad de
Valencia-CSIC Institutos de Investigaci\'on de Paterna, Aptdo.22085,
46071 Valencia, Spain}

\begin{abstract}
We have studied the meson-baryon interaction in coupled channels with the same quantum numbers of $\Xi_{bc}$.
The interaction is attractive in some channels and of sufficient intensity to lead to bound states or resonances.
We use a model describing the meson-baryon interaction based on an extrapolation of the local hidden gauge approach to the heavy sector,
which has been successfully used in predicting $\Omega_c$ and hidden charm states. We obtain many states,
some of them narrow or with zero width, as a consequence of the interaction,
which qualify as molecular states in those channels.
The success in related sectors of the picture used should encourage the experimental search for such states.
\end{abstract}

\maketitle


\section{Introduction}
The spectroscopy of baryons with heavy quarks has raised a wave of intensive theoretical work,
with models competing to explain experimental facts and make new predictions.
The reporting of several pentaquark states in Ref.~\cite{Aaij:2015tga},
updated recently \cite{Aaij:2019vzc}, was a main trigger of this wave of works,
but other works, as the discovery of several narrow $\Omega_c$ states \cite{Aaij:2017nav},
and the more recent finding of a $\Xi_{cc}^{++}$ state \cite{Aaij:2017ueg},
have also contributed to keep the flame alive.

The $\Xi_{cc}^{++}$ discovery was again a turning point,
since previous theoretical works had made predictions for its existence \cite{DeRujula:1975qlm,Ebert:2002ig,Korner:1994nh}
(see further information in Ref.~\cite{Chen:2016spr}).
After the experimental discovery \cite{Aaij:2017ueg} the attention to doubly heavy baryons experienced a boom
and those states have been considered from various points of view.
Much of the attention has been given to weak decays of these states \cite{Yu:2017zst,Wang:2017mqp,Wang:2017azm,Gutsche:2017hux,Sharma:2017txj,Hu:2017dzi,Shi:2017dto,Karliner:2018hos,Zhao:2018mrg,
Xing:2018lre,Wang:2018lhz,Cheng:2018mwu,Jiang:2018oak,Zhang:2018llc,Gutsche:2018msz,Ridgway:2019zks,Cheng:2019sxr,Gerasimov:2019jwp},
but strong and electromagnetic decays also received some attention \cite{Li:2017pxa,Xiao:2017udy,Mehen:2017nrh,Cui:2017udv,Xiao:2017dly,Bahtiyar:2018vub}.
Magnetic moments of these states have also been evaluated in different approaches \cite{Li:2017cfz,Meng:2017dni,Ozdem:2018uue,Ozdem:2019zis}.
Concerning masses and spectra of excited states, sum rules have contributed their share,
with the customary large uncertainties \cite{Wang:2017qvg,Wang:2018ihk,Azizi:2018dva,Wang:2018lhz,Aliev:2019lvd},
and have also been used to evaluate weak decays \cite{Shi:2019hbf,Shi:2019fph}.
Lattice QCD calculations have also been done for the ground states \cite{Mathur:2018rwu}.
As usual, quark models have been also used, mostly from the conventional $QQq$ structure,
to obtain spectra of doubly heavy baryons \cite{Kiselev:2017eic,Lu:2017meb,Shah:2017jkr,Zhou:2018pcv,Weng:2018mmf,Richard:2018yrm,Li:2019ekr,Albertus:2006ya}.
Detailed spectra for $\Xi_{cc}$ and $\Xi_{bc}$ states, among others, are obtained in Ref.~\cite{Shah:2017liu} using the hypercentral constituent quark model.
Heavy quark spin symmetry has also been one of the elements used to obtain spectra of doubly heavy baryons relating the different heavy flavor sectors \cite{Ma:2017nik,Liu:2018bkx,Mehen:2019cxn}.

Related works include the study of systems of a light pseudoscalar with doubly heavy baryons \cite{Guo:2017vcf,Yan:2018zdt,Meng:2018zbl},
studies of triple charm molecular pentaquarks of $\Xi_{cc}D^{(\ast)}$ systems with pion exchange \cite{Chen:2017jjn},
the use of Lorentz-invariant baryon chiral perturbation theory to study the ground state of $\Xi_{cc}$ \cite{Yao:2018ifh},
the study of electromagnetic form factors of $\Xi_{cc}$, $\Omega_{cc}$ \cite{Blin:2018pmj},
and the study of $\Xi_{cc}$, $\Xi_{bc}$ and $\Xi_{bc}^\prime$ masses using a scalar confining potential
and one gluon exchange with the Bethe-Salpeter equation \cite{qixin}.
Reviews on these topics can be found in Refs.~\cite{Olsen:2017bmm,Karliner:2017qhf,Liu:2019zoy}.
Related to the works of Refs.~\cite{Guo:2017vcf,Yan:2018zdt,Meng:2018zbl} is the work of Ref.~\cite{Dias:2018qhp},
but with more coupled channels. For instance, in addition to $\Xi_{cc}\pi$, $\Xi_{cc}\eta$, $\Omega_{cc}K$,
that account for a light pseudoscalar and a heavy baryon,
the $\Lambda_cD$, $\Sigma_cD$, $\Xi_cD_s$, $\Xi_c^\prime D_s$ channels are considered to produce $\Xi_{cc}$ states in $J^P=1/2^-$,
which by themselves give rise to molecular states.
Similar states using vector-baryon interaction and mesons with $3/2^+$ baryons are also considered in Ref.~\cite{Dias:2018qhp}.
The approach predicts several states of negative parity between $3837\,\rm MeV$ and $4374\,\rm MeV$.
It is clear that when including coupled channels with charmed mesons one can no longer invoke chiral dynamics,
as is the case when dealing with light pseudoscalar in Refs.~\cite{Guo:2017vcf,Yan:2018zdt,Meng:2018zbl}.
Instead, a method was found in Refs.~\cite{Debastiani:2017ewu,Sakai:2017avl} to produce a reliable source of interaction in this case:
\begin{itemize}
  \item [i)] First one realizes that the chiral Lagrangians in SU(3) can be obtained from the local hidden gauge approach \cite{Bando:1984ej,Bando:1987br,Meissner:1987ge,Nagahiro:2008cv} by exchanging vector mesons.
      This was shown in the case of the pseudoscalar-pseudoscalar interaction in Ref.~\cite{Ecker:1989yg}.
  \item [ii)] Take a typical channel $\Xi_cD$ and the direct transition $\Xi_cD\to\Xi_cD$.
  The $D^+$ flavor wave function is just $c\bar d$,
  and for the $\Xi_c$ and other baryons we single out the heavy quark and impose flavor-spin symmetry in the remaining light quarks.
  Thus, explicitly one is not making use of SU(4) symmetry. The direct $\Xi_cD\to\Xi_cD$ transition is mediated by the exchange of light vectors
  and $c$ quarks in $D$ and $\Xi_c$ act as spectators. Then the interaction follows the SU(3) symmetry of the light quarks. A welcome side effect is that,
  since the heavy quarks are spectators, the interaction does not depend upon them and heavy quark symmetry is automatically implemented.
  \item [iii)] Some non-diagonal transitions,
  like $\Sigma_cD\to\Xi_{cc}\pi$, require the exchange of a $D^\ast$ and here the heavy quarks are no longer spectators.
  Yet, no SU(4) is used in the approach with the wave functions used, and the vertices $VPP$, $VBB$ ($V$ for vector,
  $P$ for pseudoscalar and $B$ for baryon) essentially count the number of quarks involved in the exchange.
  Yet, these terms no longer comply with heavy quark symmetry,
      as one finds explicitly, but neither should they,
      since the interaction goes as $\mathcal O(\frac{1}{m_Q^2})$ ($m_Q$ for the mass of the heavy quark) from the $D^\ast$ propagator,
      and these terms are subleading in the $\mathcal O(\frac{1}{m_Q})$ counting, and small in practice.
  \item [iv)] One needs a piece of experimental information to fine tune the regulator of the loops (usually $q_{\rm max}$ for the modulus of the three momentum),
  which is adjusted to some mass and should be of natural size in the range of $600-800\,\rm MeV$.
      Then the masses and widths of many states are predicted by the approach.
\end{itemize}

This said it is not surprising that the approaches
which use explicitly SU(4) for the evaluation of this interaction obtain the same results
for the transition matrix elements led by the exchange of light vectors,
since automatically they are effectively using the SU(3) subgroup of this group.
This is the case of the work of Ref.~\cite{Montana:2017kjw} in the study of $\Omega_c$ molecular states.
There are differences with respect to Ref.~\cite{Debastiani:2017ewu} in the transitions including the exchange of heavy vectors,
but since this interaction is small, it is not surprising that in the end the results of Ref.~\cite{Montana:2017kjw} using SU(4)
and those of Ref.~\cite{Debastiani:2017ewu} where SU(4) is not used, are very similar.

The approach described above has been very successful,
and in Ref.~\cite{Debastiani:2017ewu} three of the $\Omega_c$ states of Ref.~\cite{Aaij:2017nav} were correctly reproduced in mass and width.
In Ref.~\cite{Xiao:2019aya} heavy quark spin symmetry (HQSS) was used
to find the relationship between the transition matrix elements of the coupled channels, $\bar D^{(\ast)}\Sigma_c^{(\ast)}$, $J/\psi N$ and others,
to describe the recent hidden charm pentaquark states \cite{Aaij:2019vzc}.
The strength of the interaction was obtained from the evaluation of the hidden gauge approach described above. Once again,
the states found in the experiment were fairly well reproduced and a few more states were predicted.
These results are similar to those of Ref.~\cite{Liu:2019tjn}, where also HQSS is used to evaluate masses,
but in addition the widths are evaluated in Ref.~\cite{Debastiani:2017ewu}.

The approach of Ref.~\cite{Debastiani:2017ewu} to study the $\Omega_c$ states is also used in Ref.~\cite{Dias:2018qhp} in the study of the $\Xi_{cc}$ states,
in Ref.~\cite{Yu:2018yxl} in the study of $\Xi_c$ and $\Xi_b$ states and in Ref.~\cite{Liang:2017ejq} in the study of $\Omega_b$ states.

In the present work we study in detail the $\Xi_{bc}$ states that can emerge from the interaction of pseudoscalar-baryon$(1/2^+)$ interaction,
vector-baryon$(1/2^+)$ interaction and pseudoscalar-baryon$(3/2^+)$ interaction.
Using the same regulator obtained in cases where we could contrast with experiment, and the same source of information,
we obtain several states in each sector. We evaluate binding energies and widths,
as well as couplings of the resonant states to the different channels.
In some cases we can see a striking dominance of one of the channels,
which allows us to deem the state as a molecular state of this channel.
Since we work in meson-baryon interaction in $s$-wave, we also evaluate the wave functions at the origin for the different channels,
which provide extra information concerning the relevance of the channels in different reactions.

\section{Formalism}
\subsection{Baryon states}
In order to see the coupled channels that we need, we classify the meson-baryon states as
\begin{itemize}
  \item [1)] Meson-baryon states with both $b$ and $c$ quarks in the baryon. For this case,
  we have baryons: $\Xi_{bc}\equiv bcq$ ($q$ for $u$ or $d$ quark); $\Omega_{bc}\equiv bcs$.
  The coupled channels of pseudoscalar meson and baryon are
      \begin{equation}
      \pi\Xi_{bc},\quad \eta\Xi_{bc},\quad K\Omega_{bc};
      \end{equation}
  \item [2)] Meson-baryon states with $b$ in the baryon and $c$ in the meson. The coupled channels are
  \begin{equation}
  D\Lambda_b,\quad D\Sigma_b,\quad D_s\Xi_b,\quad D_s\Xi_b^\prime;
  \end{equation}
  \item[3)] Meson-baryon states with $c$ in the baryon and $b$ in the meson. The coupled channels are
  \begin{equation}
  \bar B\Lambda_c,\quad \bar B\Sigma_c,\quad \bar B_s\Xi_c,\quad \bar B_s\Xi_c^\prime.
  \end{equation}
\end{itemize}

Next we take the baryon wave functions isolating the heavy quarks and imposing the spin-flavor symmetry on the light quarks.
In our approach it is important to specify the spin of the states because, as we shall see below,
the interaction is spin independent,
which immediately imposes selection rules in the transitions.
The wave functions are summarized in Tables~\ref{tab:baryonwave} and \ref{tab:baryonwave2}.

\begin{table}[H]
\renewcommand\arraystretch{1.0}
\centering
\caption{\vadjust{\vspace{-2pt}}Wave functions for baryons with $J^P=1/2^+$ and $I=0,1/2,1$.
$MS$ and $MA$ stand for mixed symmetric and mixed antisymmetric, respectively.}\label{tab:baryonwave}
\begin{tabular*}{1.00\textwidth}{@{\extracolsep{\fill}}cccc}
\hline
\hline
States         & $I,J$     & Flavor                       & Spin   \\
\hline
$\Omega_{bc}^0$& $0,1/2$   & $bcs$                        & $\chi_{MS},\chi_{MA}$\\
$\Lambda_b^0$  & $0,1/2$   & $\frac{b}{\sqrt2}(ud-du)$    & $\chi_{MA}$\\
$\Sigma_b^+$   & $1,1/2$   & $\frac{1}{\sqrt2}buu$        & $\chi_{MS}$\\
$\Xi_b^0$      & $1/2,1/2$ & $\frac{b}{\sqrt2}(us-su)$    & $\chi_{MA}$\\
$\Xi_b^{\prime0}$&$1/2,1/2$& $\frac{b}{\sqrt2}(us+su)$    & $\chi_{MS}$\\
$\Lambda_c^+$  & $0,1/2$   & $\frac{c}{\sqrt2}(ud-du)$    & $\chi_{MA}$\\
$\Sigma_c^{++}$& $1,1/2$   & $cuu$                        & $\chi_{MS}$\\
$\Xi_c^+$      & $1/2,1/2$ & $\frac{c}{\sqrt2}(us-su)$    & $\chi_{MA}$\\
$\Xi_c^{\prime0}$&$1/2,1/2$& $\frac{c}{\sqrt2}(us+su)$    & $\chi_{MS}$\\
\hline
\hline
\end{tabular*}
\end{table}
\begin{table}[H]
\renewcommand\arraystretch{1.0}
\centering
\caption{\vadjust{\vspace{-2pt}}Wave functions for baryons with $J^P=3/2^+$ and $I=0,1/2,1$.
$S$ in $\chi_S$ stands for full symmetric.}\label{tab:baryonwave2}
\begin{tabular*}{1.00\textwidth}{@{\extracolsep{\fill}}cccc}
\hline
\hline
States         & $I,J$     & Flavor                       & Spin   \\
\hline
$\Omega_{bc}^{\ast0}$& $0,3/2$   & $bcs$                  & $\chi_{S}$\\
$\Sigma_b^{\ast+}$   & $1,3/2$   & $buu$                  & $\chi_{S}$\\
$\Xi_b^{\ast0}$      & $1/2,3/2$ & $\frac{b}{\sqrt2}(us+su)$    & $\chi_{S}$\\
$\Sigma_c^{\ast++}$& $1,3/2$   & $cuu$                        & $\chi_{S}$\\
$\Xi_c^{\ast+}$      & $1/2,3/2$ & $\frac{c}{\sqrt2}(us+su)$    & $\chi_{S}$\\
\hline
\hline
\end{tabular*}
\end{table}

The corresponding states with different charge are trivial using the $u,d$ quarks.
In Tables~\ref{tab:baryonwave} and \ref{tab:baryonwave2} $\chi_{MS}$, $\chi_{MA}$,
$\chi_S$ are the spin wave functions of the three quarks \cite{Close:1979bt}.

\begin{figure}[H]
  \centering
  \includegraphics[width=0.35\textwidth]{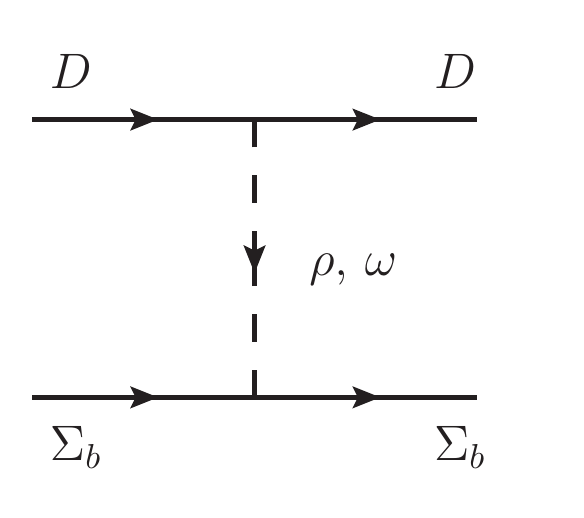}
  \caption{Example of interaction for one of the channels.}
  \label{feyn1}
\end{figure}

In the interaction we exchange vector mesons as shown in Fig.~\ref{feyn1}.
The lower vertex of $VBB$ is of the type $\gamma^\mu\epsilon_\mu$,
but, with heavy baryons and close to threshold, only the $\gamma^0\simeq1$ term is relevant, which
means that this vertex is spin independent, and so is the upper vertex that will go as $(p_D+p_D^\prime)^0$.
For $VB \to VB$ transitions, the upper vertex has the same structure,
but with the additional $\vec \epsilon \cdot \vec \epsilon\,'$ factor for the vector polarizations,
which is diagonal in the spin of the vectors.
With the spin-independent interaction, we can classify the meson-baryon channels according to the spin wave functions of baryons,
$i.e.$, $\chi_{MS}$, $\chi_{MA}$ or $\chi_S$. Hence we have the blocks of coupled channels:
\begin{itemize}
  \item [A)] $PB$ channels with $\chi_{MS}$ for the baryon:
  $\pi\Xi_{bc}$, $\eta\Xi_{bc}$, $K\Omega_{bc}$, $D\Sigma_b$, $D_s\Xi_b^\prime$, $\bar B\Sigma_c$, $\bar B_s\Xi_c^\prime$.
  \item [B)] $PB$ channels with $\chi_{MA}$ for the baryon:
  $\pi\Xi_{bc}$, $\eta\Xi_{bc}$, $K\Omega_{bc}$, $D\Lambda_b$, $D_s\Xi_b$, $\bar B\Lambda_c$, $\bar B_s\Xi_c$.
  \item [C)] $PB$ channels with $\chi_{S}$ for the baryon: $\pi\Xi_{bc}^\ast$, $\eta\Xi_{bc}^\ast$,
  $K\Omega_{bc}^\ast$, $D\Sigma_b^\ast$, $D_s\Xi_b^\ast$, $\bar B\Sigma_c^\ast$, $\bar B_s\Xi_c^\ast$.
  \item [D)] $VB$ channels with $\chi_{MS}$ for the baryon: $\rho\Xi_{bc}$, $\omega\Xi_{bc}$,
  $\phi\Xi_{bc}$, $K^\ast\Omega_{bc}$, $D^\ast\Sigma_b$, $D_s^\ast\Xi_b^\prime$, $\bar B^\ast\Sigma_c$, $\bar B_s^\ast\Xi_c^\prime$.
  \item [E)] $VB$ channels with $\chi_{MA}$ for the baryon:
  $\rho\Xi_{bc}$, $\omega\Xi_{bc}$, $\phi\Xi_{bc}$, $K^\ast\Omega_{bc}$, $D^\ast\Lambda_b$, $D_s^\ast\Xi_b$, $\bar B^\ast\Lambda_c$, $\bar B_s^\ast\Xi_c$.
\end{itemize}

We do not study the interaction of vectors with $J^P=3/2^+$ baryons.
From previous works the states obtained belong to a region where signals are difficult to see experimentally.

\subsection{Isospin states}
We take the isospin multiplets:
\par\vspace{0.5cm}
\begin{minipage}{0.40\linewidth}
\begin{equation}
\Xi_{bc} = \begin{pmatrix}
\Xi_{bc}^+ \\
\Xi_{bc}^0
\end{pmatrix},
\end{equation}
\begin{equation}
D = \begin{pmatrix}
D^+ \\
-D^0
\end{pmatrix},
\end{equation}
\begin{equation}
\bar D = \begin{pmatrix}
\bar D^0 \\
D^-
\end{pmatrix},
\end{equation}
\begin{equation}
\Xi_b = \begin{pmatrix}
\Xi_b^0 \\
\Xi_b^-
\end{pmatrix},
\end{equation}
\begin{equation}
\Xi_b^\prime = \begin{pmatrix}
\Xi_b^{\prime0} \\
\Xi_b^{\prime-}
\end{pmatrix},
\end{equation}
\begin{equation}
B = \begin{pmatrix}
B^+ \\
B^0
\end{pmatrix},
\end{equation}
\begin{equation}
\bar B = \begin{pmatrix}
\bar B^0 \\
-B^-
\end{pmatrix},
\end{equation}
\end{minipage}
\begin{minipage}{0.45\linewidth}
\begin{equation}
K = \begin{pmatrix}
K^+ \\
K^0
\end{pmatrix},
\end{equation}
\begin{equation}
\bar{K} = \begin{pmatrix}
\bar{K}^0 \\
-K^-
\end{pmatrix},
\end{equation}
\begin{equation}
\pi = \begin{pmatrix}
-\pi^+ \\
\pi^0 \\
\pi^-
\end{pmatrix},
\end{equation}
\begin{equation}
\rho = \begin{pmatrix}
-\rho^+ \\
\rho^0 \\
\rho^-
\end{pmatrix},
\end{equation}
\begin{equation}
\Sigma_b = \begin{pmatrix}
\Sigma_b^+ \\
\Sigma_b^0 \\
\Sigma_b^-
\end{pmatrix}.
\end{equation}
\end{minipage}\vspace{0.5cm}

We find the following states classified as $BP$, $BV$.
For consistency with Ref.~\cite{Dias:2018qhp} we write the states as baryon-meson. Then
\begin{align}
\big|\Sigma_b D;1/2,1/2\big>=&-\sqrt{\frac{2}{3}}\big|\Sigma_b^+D^0 \big>-\sqrt{\frac{1}{3}}\big|\Sigma_b^0D^+\big>,\\
\big|\Sigma_c\bar B;1/2,1/2\big>=&-\sqrt{\frac{2}{3}}\big|\Sigma_c^{++}B^-\big>-\sqrt{\frac{1}{3}}\big|\Sigma_c^+\bar B^0\big>,\\
\big|\Sigma_b^\ast D;1/2,1/2\big>=&-\sqrt{\frac{2}{3}}\big|\Sigma_b^{\ast+}D^0 \big>-\sqrt{\frac{1}{3}}\big|\Sigma_b^{\ast0}D^+\big>,\\
\big|\Sigma_c^\ast\bar B;1/2,1/2\big>=&-\sqrt{\frac{2}{3}}\big|\Sigma_c^{\ast++}B^- \big>-\sqrt{\frac{1}{3}}\big|\Sigma_c^{\ast+}B^0\big>,
\end{align}
and the rest are trivial.
\subsection{Evaluation of matrix elements}
The evaluation of the upper vertex of Fig.~\ref{feyn1}, $VPP$,
is done in Ref.~\cite{Sakai:2017avl} using the vector character of the vertex and the quark content of the mesons.
Yet, it was found that for practical purpose one can get it from the Lagrangian
\begin{equation}
\mathcal L=-ig\langle\left[P,\partial_\mu P\right]V^\mu\rangle,
\end{equation}
where $\langle\cdots\rangle$ stands for the matrix trace, $g=M_V/2f_\pi$, $M_V=800\,\rm MeV$, $f_\pi=93\,\rm MeV$ and
\begin{equation}
P = \begin{pmatrix}
\frac{1}{\sqrt{2}}\pi^0 + \frac{1}{\sqrt{3}} \eta + \frac{1}{\sqrt{6}}\eta' & \pi^+ & K^+ & \bar{D}^0 \\
 \pi^- & -\frac{1}{\sqrt{2}}\pi^0 + \frac{1}{\sqrt{3}} \eta + \frac{1}{\sqrt{6}}\eta' & K^0 & D^- \\
 K^- & \bar{K}^0 & -\frac{1}{\sqrt{3}} \eta + \sqrt{\frac{2}{3}}\eta' & D_s^- \\
D^0  & D^+ & D_s^+ & \eta_c
\end{pmatrix},
\end{equation}
\begin{equation}
\label{eq:matV}
V = \begin{pmatrix}
 \frac{1}{\sqrt{2}}\rho^0 + \frac{1}{\sqrt{2}} \omega & \rho^+ & K^{* +} & \bar{D}^{* 0} \\
 \rho^- & -\frac{1}{\sqrt{2}}\rho^0 + \frac{1}{\sqrt{2}} \omega  & K^{* 0} & \bar{D}^{* -} \\
 K^{* -} & \bar{K}^{* 0}  & \phi & D_s^{* -} \\
 D^{* 0} & D^{* +} & D_s^{* +} & J/\psi
\end{pmatrix},
\end{equation}
when dealing with charmed mesons, and
\begin{equation}
P = \begin{pmatrix}
\frac{1}{\sqrt{2}}\pi^0 + \frac{1}{\sqrt{3}} \eta + \frac{1}{\sqrt{6}}\eta' & \pi^+ & K^+ & B^+ \\
 \pi^- & -\frac{1}{\sqrt{2}}\pi^0 + \frac{1}{\sqrt{3}} \eta + \frac{1}{\sqrt{6}}\eta' & K^0 & B^0 \\
 K^- & \bar{K}^0 & -\frac{1}{\sqrt{3}} \eta + \sqrt{\frac{2}{3}}\eta' & B_s^0 \\
B^-  & \bar B^0 & \bar B_s^0 & \eta_b
\end{pmatrix},
\end{equation}
\begin{equation}
\label{eq:matV}
V = \begin{pmatrix}
 \frac{1}{\sqrt{2}}\rho^0 + \frac{1}{\sqrt{2}} \omega & \rho^+ & K^{* +} & B^{* +} \\
 \rho^- & -\frac{1}{\sqrt{2}}\rho^0 + \frac{1}{\sqrt{2}} \omega  & K^{* 0} & B^{* 0} \\
 K^{* -} & \bar{K}^{* 0}  & \phi & B_s^{* 0} \\
 B^{* -} & \bar B^{* 0} & \bar B_s^{* 0} & \Upsilon
\end{pmatrix},
\end{equation}
when dealing with bottomed mesons.

The lower vertex is trivial to evaluate.
The Lagrangian for the $VBB$ vertex is given by the operator
\begin{equation}
\mathcal L\rightarrow \left\{\begin{matrix}&\displaystyle\frac{g}{\sqrt2}(u\bar u-d\bar d),\,\,\text{for}\,\,\rho^0\\
                                      &\displaystyle\frac{g}{\sqrt2}(u\bar u+d\bar d),\,\,\text{for}\,\,\omega\end{matrix}\right.
\end{equation}

Hence, for instance, $\rho^0\Xi_{bc}^0\Xi_{bc}^0$ involves the vertex
\begin{equation}
\left<bcd\Big|\frac{g}{\sqrt2}(u\bar u-d\bar d)\Big|bcd\right>=-g\frac{1}{\sqrt2}.
\end{equation}

Then we get finally transition matrix elements\
\begin{equation}\label{eq:kernelnr}
V_{ij} = D_{ij} \frac{1}{4 f_{\pi}^2} (k^0+k'^0),
\end{equation}
where $k^0,k'^0$ are the energies of the mesons and $D_{ij}$ the coefficients which are shown in Tables in the next section.
With the potential of Eq.~\eqref{eq:kernelnr} we solve the Bethe-Salpeter equation in coupled channels
\begin{equation}\label{eq:BS}
T=[1-VG]^{-1}V,
\end{equation}
where $G$ is the meson-baryon loop function
\begin{align}\label{loopfunc}
  G_l=&i\int\frac{d^4q}{(2\pi)^4}\frac{M_l}{E_l(\textbf{q})}\frac{1}{k^0+p^0-q^0-E_l(\textbf{q})+i\epsilon}\frac{1}{q^2-m_l^2+i\epsilon}\nonumber\\
     =&\int_{|\textbf{q}|<q_{\rm max}}\frac{d^3\textbf{q}}{(2\pi)^3}\frac{1}{2\omega_l(\textbf{q})}
     \frac{M_l}{E_l(\textbf{q})}\frac{1}{k^0+p^0-\omega_l(\textbf{q})-E_l(\textbf{q})+i\epsilon},
\end{align}
and we use $q_{\rm max}=650\,\rm MeV$ as in Refs.~\cite{Dias:2018qhp,Debastiani:2017ewu,Liang:2017ejq}.
$\omega_l$, $E_l$ are the energies of the meson and baryon respectively, $\omega_l=\sqrt{m_l^2+\textbf{q}^2}$, $E_l=\sqrt{M_l^2+\textbf{q}^2}$,
and $m_l$, $M_l$ the meson and baryon masses. In Eq.~\eqref{loopfunc} $p^0$ is the energy of the incoming baryon.

In order to get poles in the second Riemann sheet we replace $G$ by $G^{II}$, and it is given by
\begin{equation}\label{second loop}
G_l^{\uppercase\expandafter{\romannumeral2}}= \left\{\begin{matrix}
G_l(s) & \text{for } \text{Re}(\sqrt{s})<\sqrt{s}_{th, l}\\
G_l(s)+i\displaystyle\frac{2M_lq}{4\pi\sqrt s} & \text{for } \text{Re}(\sqrt{s})\geq  \sqrt{s}_{th, l}
\end{matrix}\right.,
\end{equation}
where $\sqrt{s}_{th,l}$ is the threshold mass of the $l$ channel, and
\begin{equation}\label{momentum}
q=\frac{\lambda^{1/2}(s,m_l^2,M_l^2)}{2\sqrt s} \quad \text{with Im(q)$>$0}.
\end{equation}
For the evaluation of the couplings $g_l$ of the state to different coupled channels,
we find that the $T_{ij}$ matrix can be expressed in the following form close to the poles, $z_R$,
\begin{equation}\label{coupling1}
T_{ij}(s)=\frac{g_ig_j}{\sqrt s-z_R},
\end{equation}
which defines the couplings $g_i$ up to a global sign of one of them.

The approach followed relies on the exchange of vector mesons,
which we justified from the extension of the chiral Lagrangians via the local hidden gauge approach.
In meson-baryon interactions the exchange of pseudoscalar mesons is also sometimes used \cite{slzhu},
together with $\sigma$-exchange.
Unlike in baryon-baryon interaction where the $\pi$ exchange plays a very important role,
in meson-meson or meson-baryon interaction, pion exchange plays a more moderate role because the direct $PB\to PB$ through $\pi$ exchange does not go,
since a three-pseudoscalar vertex is forbidden by parity-angular-momentum conservation.
It is through intermediate states transitions $PB\to VB\to PB$ that the $\pi$-exchange can have a contribution.
Detailed calculations of its effects have also been considered in the study of baryon states with open charm \cite{uchino}
and hidden charm \cite{uchidos} and the main source of interaction remains vector exchange.
Also, to some extend,
the two step process $PB\to VB\to PB$ can be incorporated into an effective transition potential $\delta V$ for $PB\to PB$,
adding to the vector exchange potential,
and this extra potential can again be effectively accounted for by changes in the cut off that regularizes the loop function $G$,
since $[V^{-1}-G]$ will be the same with an extra $\delta V$ and $\delta G = \delta (V^{-1})$.
Yet, not all the contributions can be reabsorbed in this way
and one may expect some remnant contributions to break the degeneracy between spin-parity $1/2^-$
and $3/2^-$.
That these effects are finally small can be seen in the breakup of the degeneracy
of the $P_c(4440)$ and $P_c(4457)$ states recently observed in Ref.~\cite{Aaij:2019vzc}.
There are several works including pion exchange that break the degeneracy of the $1/2^-, 3/2^-$ states around this energy \cite{hepen,xliupen,Pavon,genpen},
but the small difference of mass between these states gives us an idea of the role played by $\pi$ exchange in these cases.
Uncertainties of this type in our predictions must certainly be admitted.

As to $\sigma$-exchange, it is empirically accounted for in some of the former works,
but the strength is unknown.
However, there is one way to make this exchange quantitative by going to the root
of the $\sigma$ meson as dynamically generated from the $\pi\pi$ interaction \cite{npa,sigma}.
Indeed, in Ref.~\cite{mizobe} the $\sigma$-exchange between nucleons was studied from the point of view of the exchange of two interacting mesons in $s$-wave.
The same picture was used in the study of the meson-meson interaction describing the $Z_c(3980)$ \cite{Acetione} and $Z_c(4000)$ \cite{Acetidos}
and its effect was found minor.
In particular, for meson-baryon interaction, the $\sigma$-exchange from this perspective was considered in Ref.~\cite{albersig}
in the study of the $\bar K N$ interaction and it was found very much suppressed.

\section{Results}
\subsection{Pseudoscalar-baryon$(1/2^+)$ states, mixed symmetric sector}
In Table~\ref{tab:pbms},
we consider the channels in this sector together with their threshold masses\footnote{The masses of the states not reported in the PDG are taken from the quark model calculation of Ref.~\cite{Albertus:2006ya}.}.
\begin{table}[H]
\renewcommand\arraystretch{1.0}
\begin{adjustwidth}{-0.00\textwidth}{-0.00\textwidth}
\caption{\vadjust{\vspace{-2pt}}Channels considered for sector $J^P=1/2^-$ (MS).}\label{tab:pbms}
\begin{tabular*}{1.00\textwidth}{@{\extracolsep{\fill}}c|ccccccc}
\hline
\hline
\textbf{Channel} & $\Xi_{bc}\pi$ & $\Xi_{bc}\eta$ & $\Omega_{bc}K$ & $\Sigma_bD$ & $\Sigma_c \bar B$ & $\Xi_b^\prime D_s$ & $\Xi_c^\prime \bar B_s$  \\
\hline
\textbf{Threshold (MeV)} & 7057 & 7467 & 7482 & 7680 & 7733 & 7903 & 7945\\
\hline
\hline
\end{tabular*}
\end{adjustwidth}
\end{table}
\begin{table}[H]
\renewcommand\arraystretch{1.00}
\begin{adjustwidth}{-0.00\textwidth}{-0.0\textwidth}
\caption{\vadjust{\vspace{-2pt}}$D_{ij}$ coefficients for sector $J^P=1/2^-$ (MS).}\label{PBMS}
\begin{tabular*}{1.00\textwidth}{@{\extracolsep{\fill}}c|ccccccc}
\hline
\hline
\textbf{$J^P=1/2^-$} & $\Xi_{bc}\pi$ & $\Xi_{bc}\eta$ & $\Omega_{bc}K$ & $\Sigma_bD$ & $\Sigma_c \bar B$ & $\Xi_b^\prime D_s$ & $\Xi_c^\prime \bar B_s$ \\
\hline
$\Xi_{bc}\pi$ & $-2$ & 0 & $\sqrt\frac{3}{2}$ & $\frac{1}{2}\lambda$ & 0 & 0 & 0 \\
\hline
$\Xi_{bc}\eta$ &  & 0 & $-\frac{2}{\sqrt 3}$ & $-\frac{1}{\sqrt 2}\lambda$ & 0 & $-\frac{1}{\sqrt 6}\lambda$ & 0 \\
\hline
$\Omega_{bc}K$ &  &  & $-1$ & 0 & 0 & $\frac{1}{\sqrt 2}\lambda$ & 0\\
\hline
$\Sigma_b D$ &  &  &  & $-3$ & 0 & $\sqrt 3$ & 0 \\
\hline
$\Sigma_c\bar B$ &  &  &  &  & $-3$ & 0 & $\sqrt 3$ \\
\hline
$\Xi_b^\prime D_s$ &  &  &  &  &  & $-1$ & $\sim$0 \\
\hline
$\Xi_c^\prime\bar B_s$ &  &  &  &  &  &  & $-1$ \\
\hline
\hline
\end{tabular*}
\end{adjustwidth}
\end{table}
In Table~\ref{PBMS} we show the coefficients $D_{ij}$ of Eq.~\eqref{eq:kernelnr}. We find that there are some terms which go with $\lambda$.
These terms involve the exchange of $D^\ast$ and they are reduced.
The value of $\lambda$ is the reduction factor versus the exchange of a light vector,
which in Ref.~\cite{Debastiani:2017ewu} is found to be around $\lambda=0.25$ as in Ref.~\cite{Mizutani:2006vq}.
Apart from that, when exchanging $B_c$ meson we set the $D_{ij}$ result as $\sim$0, since it is highly suppressed.

In Table~\ref{pbmspole} the poles for three states that appear in this sector are shown for different values of the cutoff $q_{\rm max}$.
As in former works we take $q_{\rm max}=650\,\rm MeV$.
In Table~\ref{pbmspole1} and \ref{pbmspole2} we take the first two states of Table~\ref{pbmspole} and show the couplings $g_i$,
and the wave functions at the origin $g_iG_i$ (see Ref.~\cite{Gamermann:2009uq}).
We see that the first state at $7131\,\rm MeV$ has a large width of about $200\,\rm MeV$ and couples mostly to $\Xi_{bc}\pi$.
However, the second state at $7372\,\rm MeV$ has a very narrow width and is basically a $\Sigma_b D$ bound state,
although it also couples strongly to $\Xi_b^\prime D_s$. The last state in Table~\ref{pbmspole} for $q_{\rm max}=650\,\rm MeV$,
appears at the same energy as the former one but the couplings are different as can be seen in Table~\ref{pbmspole3}.
\begin{table}[H]
\renewcommand\arraystretch{1.0}
\centering
\caption{\vadjust{\vspace{-2pt}}Poles for pesudoscalar-baryon(1/2) (MS) states (all units are in MeV).}\label{pbmspole}
\begin{tabular*}{1.00\textwidth}{@{\extracolsep{\fill}}ccccccc}
\hline
 $q_{\rm max}$    &600            &650            &700            &750            &800 \\
\hline
              &7132.04+i104.00&7131.50+i95.13 &7130.05+i86.36 &7127.88+i77.57 &7126.10+i68.25 \\
\hline
              &7434.02+i0.52  &7372.22+i0.64  &7305.50+i0.98  &7235.38+i1.72  &7162.72+i3.43 \\
\hline
              &7450.31        &7372.55        &7285.56        &7190.26        &7087.81 \\
\hline
\hline
\end{tabular*}
\end{table}

As it can be seen in Table~\ref{pbmspole}, the last two states in the row of $q_{\rm max}=650\,\rm MeV$ are less stable with the changes of $q_{\rm max}$ and,
hence, more uncertain.
\begin{table}[H]
\renewcommand\arraystretch{1.0}
\centering
\caption{\vadjust{\vspace{-2pt}}The coupling constants to various channels and $g_iG^{\uppercase\expandafter{\romannumeral2}}_i$ in MeV with $q_{\rm max}=650\,\rm MeV$.}\label{pbmspole1}
\begin{tabular*}{\textwidth}{@{\extracolsep{\fill}}ccccc}
\hline
\hline
\textbf{7131.50+i95.13}& $\Xi_{bc}\pi$ & $\Xi_{bc}\eta$ & $\Omega_{bc}K$ & $\Sigma_bD$ \\
\hline
 $g_i$                 &1.70+i1.23&-0.03-i0.10&-0.85-i0.74&-0.67-i0.40\\
 $g_iG^{\uppercase\expandafter{\romannumeral2}}_i$&-73.84-i12.90&0.05+i0.64&4.46+i5.46&1.18+i0.96\\
\hline
                      & $\Sigma_c \bar B$ & $\Xi_b^\prime D_s$ & $\Xi_c^\prime \bar B_s$ &\\
\hline
 $g_i$                &0&0.18+i0.19&0&\\
 $g_iG^{\uppercase\expandafter{\romannumeral2}}_i$&0&-0.22-i0.29&0&\\
 \hline
\end{tabular*}
\end{table}
\begin{table}[H]
\renewcommand\arraystretch{1.0}
\centering
\caption{\vadjust{\vspace{-2pt}}The coupling constants to various channels and $g_iG^{\uppercase\expandafter{\romannumeral2}}_i$ in MeV with $q_{\rm max}=650\,\rm MeV$.}\label{pbmspole2}
\begin{tabular*}{\textwidth}{@{\extracolsep{\fill}}ccccc}
\hline
\hline
\textbf{7372.22+i0.64}& $\Xi_{bc}\pi$ & $\Xi_{bc}\eta$ & $\Omega_{bc}K$ & $\Sigma_bD$ \\
\hline
 $g_i$                 &-0.02+i0.14&0.27-i0.03&0.08-i0.10&9.40-i0.02\\
 $g_iG^{\uppercase\expandafter{\romannumeral2}}_i$&-4.27-i2.06&-3.33+i0.38&-0.95+i1.14&-29.42+i0.01\\
\hline
                      & $\Sigma_c \bar B$ & $\Xi_b^\prime D_s$ & $\Xi_c^\prime \bar B_s$ &\\
\hline
 $g_i$                &0&-5.19+i0.01&0&\\
 $g_iG^{\uppercase\expandafter{\romannumeral2}}_i$&0&10.02-i0.01&0&\\
 \hline
\end{tabular*}
\end{table}
\begin{table}[H]
\renewcommand\arraystretch{1.0}
\centering
\caption{\vadjust{\vspace{-2pt}}The coupling constants to various channels and $g_iG^{\uppercase\expandafter{\romannumeral2}}_i$ in MeV with $q_{\rm max}=650\,\rm MeV$.}\label{pbmspole3}
\begin{tabular*}{\textwidth}{@{\extracolsep{\fill}}ccccc}
\hline
\hline
\textbf{7372.55}& $\Xi_{bc}\pi$ & $\Xi_{bc}\eta$ & $\Omega_{bc}K$ & $\Sigma_bD$ \\
\hline
 $g_i$                 &0&0&0&0\\
 $g_iG^{\uppercase\expandafter{\romannumeral2}}_i$&0&0&0&0\\
\hline
                      & $\Sigma_c \bar B$ & $\Xi_b^\prime D_s$ & $\Xi_c^\prime \bar B_s$ &\\
\hline
 $g_i$                &18.10&0&-10.17&\\
 $g_iG^{\uppercase\expandafter{\romannumeral2}}_i$&-18.43&0&6.96&\\
 \hline
\end{tabular*}
\end{table}
\subsection{Pseudoscalar-baryon$(1/2^+)$ states, mixed antisymmetric sector}
We follow here the same pattern as in the former subsection with the mixed antisymmetric states.
In Table~\ref{tab:pbma} we show the channels and the threshold masses. In Table~\ref{PBMA} we show the $D_{ij}$ coefficients.
In Table~\ref{pbmapole} we show the states (poles) found in this sector. We find three clear states,
one with a large width of almost $200\,\rm MeV$ and two more states with very narrow width.
The first state is actually the same one that we found before, because it couples mostly to $\Xi_{bc}\pi$ as shown in Table~\ref{pbmapole1},
and this state appears with $MS$ and $MA$ spins. However, the second state couples mostly to $\Lambda_bD$ and $\Xi_bD_s$ as shown in Table~\ref{pbmapole2},
and is then a different state. The state at $7462\,\rm MeV$ couples mostly to $\Lambda_c\bar B$ and $\Xi_c\bar B_s$, as shown in Table~\ref{pbmapole3}.
\begin{table}[H]
\renewcommand\arraystretch{1.0}
\begin{adjustwidth}{-0.00\textwidth}{-0.00\textwidth}
\caption{\vadjust{\vspace{-2pt}}Channels considered for sector $J^P=1/2^-$ (MA).}\label{tab:pbma}
\begin{tabular*}{1.00\textwidth}{@{\extracolsep{\fill}}c|ccccccc}
\hline
\hline
\textbf{Channel} & $\Xi_{bc}\pi$ & $\Xi_{bc}\eta$ & $\Omega_{bc}K$ & $\Lambda_bD$ & $\Lambda_c \bar B$ & $\Xi_b D_s$ & $\Xi_c \bar B_s$  \\
\hline
\textbf{Threshold (MeV)} & 7057 & 7467 & 7482 & 7487 & 7565 & 7761 & 7836\\
\hline
\hline
\end{tabular*}
\end{adjustwidth}
\end{table}
\begin{table}[H]
\renewcommand\arraystretch{1.0}
\begin{adjustwidth}{-0.00\textwidth}{-0.0\textwidth}
\caption{\vadjust{\vspace{-2pt}}$D_{ij}$ coefficients for sector $J^P=1/2^-$ (MA).}\label{PBMA}
\begin{tabular*}{1.00\textwidth}{@{\extracolsep{\fill}}c|ccccccc}
\hline
\hline
\textbf{$J^P=1/2^-$} & $\Xi_{bc}\pi$ & $\Xi_{bc}\eta$ & $\Omega_{bc}K$ & $\Lambda_bD$ & $\Lambda_c \bar B$ & $\Xi_b D_s$ & $\Xi_c \bar B_s$  \\
\hline
$\Xi_{bc}\pi$ & $-2$ & 0 & $\sqrt\frac{3}{2}$ & $-\frac{\sqrt 3}{2}\lambda$ & 0 & 0 & 0 \\
\hline
$\Xi_{bc}\eta$ &  & 0 & $-\frac{2}{\sqrt 3}$ & $-\frac{1}{\sqrt 6}\lambda$ & 0 & $\frac{1}{\sqrt 6}\lambda$ & 0 \\
\hline
$\Omega_{bc}K$ &  &  & $-1$ & 0 & 0 & $\frac{1}{\sqrt 2}\lambda$ & 0\\
\hline
$\Lambda_b D$ &  &  &  & $-1$ & $\sim$0 & $-1$ & 0 \\
\hline
$\Lambda_c\bar B$ &  &  &  &  & $-1$ & 0 & $-1$ \\
\hline
$\Xi_b D_s$ &  &  &  &  &  & $-1$ & $\sim$0 \\
\hline
$\Xi_c\bar B_s$ &  &  &  &  &  &  & $-1$ \\
\hline
\hline
\end{tabular*}
\end{adjustwidth}
\end{table}
\begin{table}[H]
\renewcommand\arraystretch{1.0}
\centering
\caption{\vadjust{\vspace{-2pt}}Poles for pesudoscalar-baryon(1/2) (MA) states (all units are in MeV).}\label{pbmapole}
\begin{tabular*}{1.02\textwidth}{@{\extracolsep{\fill}}ccccccc}
\hline
 $q_{\rm max}$    &600            &650            &700            &750            &800 \\
\hline
              &7131.20+i102.93&7130.33+i93.80 &7128.44+i84.77 &7125.60+i75.91 &7121.84+i67.11 \\
\hline
              &7428.22+i0.52  &7403.51+i0.93  &7373.53+i1.52  &7338.59+i2.35  &7299.25+i3.59 \\
\hline
              &7492.72        &7462.49        &7425.33        &7381.21        &7330.34 \\
\hline
\hline
\end{tabular*}
\end{table}
\begin{table}[H]
\renewcommand\arraystretch{1.0}
\centering
\caption{\vadjust{\vspace{-2pt}}Coupling constants to various channels and $g_iG^{\uppercase\expandafter{\romannumeral2}}_i$ in MeV with $q_{\rm max}=650\,\rm MeV$.}\label{pbmapole1}
\begin{tabular*}{\textwidth}{@{\extracolsep{\fill}}ccccc}
\hline
\hline
\textbf{7130.33+i93.80}& $\Xi_{bc}\pi$ & $\Xi_{bc}\eta$ & $\Omega_{bc}K$ & $\Lambda_bD$ \\
\hline
 $g_i$                 &1.70+i1.23&-0.01-i0.08&-0.85-i0.73&1.00+i0.42\\
 $g_iG^{\uppercase\expandafter{\romannumeral2}}_i$&-73.60-i12.80&-0.03+i0.52&4.44+i5.35&-2.45-i1.67\\
\hline
                      & $\Lambda_c \bar B$ & $\Xi_b D_s$ & $\Xi_c \bar B_s$ &\\
\hline
 $g_i$                &0&0.24+i0.23&0&\\
 $g_iG^{\uppercase\expandafter{\romannumeral2}}_i$&0&-0.35-i0.43&0&\\
 \hline
\end{tabular*}
\end{table}
\begin{table}[H]
\renewcommand\arraystretch{1.0}
\centering
\caption{\vadjust{\vspace{-2pt}}Coupling constants to various channels and $g_iG^{\uppercase\expandafter{\romannumeral2}}_i$ in MeV with $q_{\rm max}=650\,\rm MeV$.}\label{pbmapole2}
\begin{tabular*}{\textwidth}{@{\extracolsep{\fill}}ccccc}
\hline
\hline
\textbf{7403.51+i0.93}& $\Xi_{bc}\pi$ & $\Xi_{bc}\eta$ & $\Omega_{bc}K$ & $\Lambda_bD$ \\
\hline
 $g_i$                 &0.02-i0.17&0.24+i0.05&0.25+i0.13&4.06-i0.05\\
 $g_iG^{\uppercase\expandafter{\romannumeral2}}_i$&5.42+i1.92&-3.40-i0.75&-3.35-i1.72&-30.28+i0.20\\
\hline
                      & $\Lambda_c \bar B$ & $\Xi_b D_s$ & $\Xi_c \bar B_s$ &\\
\hline
 $g_i$                &0&3.85-i0.05&0&\\
 $g_iG^{\uppercase\expandafter{\romannumeral2}}_i$&0&-10.26+i0.11&0&\\
 \hline
\end{tabular*}
\end{table}
\begin{table}[H]
\renewcommand\arraystretch{1.0}
\centering
\caption{\vadjust{\vspace{-2pt}}Coupling constants to various channels and $g_iG^{\uppercase\expandafter{\romannumeral2}}_i$ in MeV with $q_{\rm max}=650\,\rm MeV$.}\label{pbmapole3}
\begin{tabular*}{\textwidth}{@{\extracolsep{\fill}}ccccc}
\hline
\hline
\textbf{7462.49}& $\Xi_{bc}\pi$ & $\Xi_{bc}\eta$ & $\Omega_{bc}K$ & $\Lambda_bD$ \\
\hline
 $g_i$                 &0&0&0&0\\
 $g_iG^{\uppercase\expandafter{\romannumeral2}}_i$&0&0&0&0\\
\hline
                      & $\Lambda_c \bar B$ & $\Xi_b D_s$ & $\Xi_c \bar B_s$ &\\
\hline
 $g_i$                &7.46&0&7.19&\\
 $g_iG^{\uppercase\expandafter{\romannumeral2}}_i$&-18.58&0&-7.01&\\
 \hline
\end{tabular*}
\end{table}

\subsection{Vector-baryon$(1/2^+)$ states, mixed symmetric sector}
We show in Table~\ref{tab:vbms} the channels and the thresholds that in this case give rise to degenerate states in sector $J^P=1/2^-,3/2^-$.
The $D_{ij}$ coefficients are shown in Table~\ref{VBMS} and in Table~\ref{vbmspole} we find four states,
three of them with zero width and the last one with a large width.
In Tables~\ref{vbmspole1}, \ref{vbmspole2} \ref{vbmspole3} and \ref{vbmspole4} we show the couplings and wave functions of these states.
The state at $7418\,\rm MeV$ could be regarded as a $\Sigma_c\bar B^\ast$ bound state,
but it also couples strongly to $\Xi_c^\prime\bar B_s^\ast$. The state at $7501\,\rm MeV$ could be identified with a $\Sigma_bD^\ast$ state,
the one at $7595\,\rm MeV$ as a $\Xi_{bc}\rho$ bound state,
and the state at $7837\,\rm MeV$ and $\Gamma\simeq180\,\rm MeV$ corresponds to a $\Omega_{bc}K^\ast$ bound state which decays strongly in $\Xi_{bc}\omega$.
\begin{table}[H]
\renewcommand\arraystretch{1.0}
\centering
\caption{\vadjust{\vspace{-2pt}}Channels considered for sector $J^P=1/2^-,3/2^-$ (MS).}\label{tab:vbms}
\begin{tabular*}{1.00\textwidth}{@{\extracolsep{\fill}}c|ccccccccc}
\hline
\hline
\textbf{Channel} & $\Xi_{bc} \rho$ & $\Xi_{bc} \omega$ &
$\Sigma_c \bar B^\ast$ & $\Sigma_b D^\ast$ & $\Omega_{bc}K^\ast$ & $\Xi_{bc}\phi$ & $\Xi_c^\prime\bar B_s^\ast$ & $\Xi_b^\prime D_s^\ast$\\
\hline
\textbf{Threshold (MeV)} & 7694 & 7702 & 7779 & 7822 & 7882 & 7938 & 7993 & 8047 \\
\hline
\hline
\end{tabular*}
\end{table}
\begin{table}[H]
\renewcommand\arraystretch{1.00}
\begin{adjustwidth}{0\textwidth}{0\textwidth}
\caption{\vadjust{\vspace{-2pt}}$D_{ij}$ coefficients for sector $J^P=1/2^-,3/2^-$ (MS).}\label{VBMS}
\begin{tabular*}{1.0\textwidth}{@{\extracolsep{\fill}}c|cccccccc}
\hline
\hline
\textbf{$J^P=1/2^-,3/2^-$} & $\Xi_{bc} \rho$ & $\Xi_{bc} \omega$ & $\Sigma_c \bar B^\ast$ &
$\Sigma_b D^\ast$ & $\Omega_{bc}K^\ast$ & $\Xi_{bc}\phi$ & $\Xi_c^\prime\bar B_s^\ast$ & $\Xi_b^\prime D_s^\ast$\\
\hline
$\Xi_{bc} \rho$ & $-2$ & 0 & 0 & $\frac{1}{2}\lambda$ & $\sqrt\frac{3}{2}$ & 0 & 0 & 0 \\
\hline
$\Xi_{bc} \omega$ &  & 0 & 0 & $-\frac{\sqrt 3}{2}\lambda$ & $-\frac{1}{\sqrt 2}$ & 0 & 0 & 0 \\
\hline
$\Sigma_c \bar B^\ast$ &  &  & $-3$ & 0 & 0 & 0 & $\sqrt 3$ & 0 \\
\hline
$\Sigma_b D^\ast$ &  &  &  & $-3$ & 0 & 0 & 0 & $\sqrt 3$ \\
\hline
$\Omega_{bc}K^\ast$ &  &  &  &  & $-1$ & 1 & 0 & $\frac{1}{\sqrt 2}\lambda$ \\
\hline
$\Xi_{bc}\phi$ &  &  &  &  &  & 0 & 0 & $\frac{1}{\sqrt 2}\lambda$ \\
\hline
$\Xi_c^\prime\bar B_s^\ast$ &  &  &  &  &  &  & $-1$ & 0 \\
\hline
$\Xi_b^\prime D_s^\ast$ &  &  &  &  &  &  &  & $-1$ \\
\hline
\hline
\end{tabular*}
\end{adjustwidth}
\end{table}
\begin{table}[H]
\renewcommand\arraystretch{1.0}
\centering
\caption{\vadjust{\vspace{-2pt}}Poles for vector-baryon(1/2) (MS) states (all units are in MeV).}\label{vbmspole}
\begin{tabular*}{1.0\textwidth}{@{\extracolsep{\fill}}ccccccc}
\hline
 $q_{\rm max}$    &600            &650            &700            &750            &800 \\
\hline
              &7496.28        &7418.47        &7331.40        &7236.01        &7133.42 \\
\hline
              &7565.71        &7501.56        &7431.85        &7358.22        &7282.17 \\
\hline
              &7619.74        &7595.24        &7569.14        &7541.98        &7514.40 \\
\hline
              &7861.84+i93.45 &7837.81+i91.45 &7810.81+86.76  &7784.37+i79.86 &7758.25+i70.40 \\
\hline
\hline
\end{tabular*}
\end{table}
\begin{table}[H]
\renewcommand\arraystretch{1.0}
\centering
\caption{\vadjust{\vspace{-2pt}}Coupling constants to various channels
and $g_iG^{\uppercase\expandafter{\romannumeral2}}_i$ in MeV with $q_{\rm max}=650\,\rm MeV$.}\label{vbmspole1}
\begin{tabular*}{\textwidth}{@{\extracolsep{\fill}}ccccc}
\hline
\hline
\textbf{7418.47}& $\Xi_{bc} \rho$ & $\Xi_{bc} \omega$ & $\Sigma_c \bar B^\ast$ & $\Sigma_b D^\ast$ \\
\hline
 $g_i$                 &0&0&18.18&0\\
 $g_iG^{\uppercase\expandafter{\romannumeral2}}_i$&0&0&-18.36&0\\
\hline
                      & $\Omega_{bc}K^\ast$ & $\Xi_{bc}\phi$ & $\Xi_c^\prime\bar B_s^\ast$ & $\Xi_b^\prime D_s^\ast$\\
\hline
 $g_i$                &0&0&-10.22&0\\
 $g_iG^{\uppercase\expandafter{\romannumeral2}}_i$&0&0&6.91&0\\
 \hline
\end{tabular*}
\end{table}
\begin{table}[H]
\renewcommand\arraystretch{1.0}
\centering
\caption{\vadjust{\vspace{-2pt}}Coupling constants to various channels and
$g_iG^{\uppercase\expandafter{\romannumeral2}}_i$ in MeV with $q_{\rm max}=650\,\rm MeV$.}\label{vbmspole2}
\begin{tabular*}{\textwidth}{@{\extracolsep{\fill}}ccccc}
\hline
\hline
\textbf{7501.56}& $\Xi_{bc} \rho$ & $\Xi_{bc} \omega$ & $\Sigma_c \bar B^\ast$ & $\Sigma_b D^\ast$ \\
\hline
 $g_i$                 &-0.62&0.44&0&9.91\\
 $g_iG^{\uppercase\expandafter{\romannumeral2}}_i$&4.76&-3.28&0&-28.37\\
\hline
                      & $\Omega_{bc}K^\ast$ & $\Xi_{bc}\phi$ & $\Xi_c^\prime\bar B_s^\ast$ & $\Xi_b^\prime D_s^\ast$\\
\hline
 $g_i$                &0.41&0.05&0&-5.47\\
 $g_iG^{\uppercase\expandafter{\romannumeral2}}_i$&-1.85&-0.18&0&9.72\\
 \hline
\end{tabular*}
\end{table}
\begin{table}[H]
\renewcommand\arraystretch{1.0}
\centering
\caption{\vadjust{\vspace{-2pt}}Coupling constants to various channels
and $g_iG^{\uppercase\expandafter{\romannumeral2}}_i$ in MeV with $q_{\rm max}=650\,\rm MeV$.}\label{vbmspole3}
\begin{tabular*}{\textwidth}{@{\extracolsep{\fill}}ccccc}
\hline
\hline
\textbf{7595.24}& $\Xi_{bc} \rho$ & $\Xi_{bc} \omega$ & $\Sigma_c \bar B^\ast$ & $\Sigma_b D^\ast$ \\
\hline
 $g_i$                 &3.58&-0.28&0&1.03\\
 $g_iG^{\uppercase\expandafter{\romannumeral2}}_i$&-38.74&2.93&0&-3.84\\
\hline
                      & $\Omega_{bc}K^\ast$ & $\Xi_{bc}\phi$ & $\Xi_c^\prime\bar B_s^\ast$ & $\Xi_b^\prime D_s^\ast$\\
\hline
 $g_i$                &-2.36&0.50&0&-0.66\\
 $g_iG^{\uppercase\expandafter{\romannumeral2}}_i$&13.01&-2.24&0&1.38\\
 \hline
\end{tabular*}
\end{table}
\begin{table}[H]
\renewcommand\arraystretch{1.0}
\centering
\caption{\vadjust{\vspace{-2pt}}Coupling constants to various channels
and $g_iG^{\uppercase\expandafter{\romannumeral2}}_i$ in MeV with $q_{\rm max}=650\,\rm MeV$.}\label{vbmspole4}
\begin{tabular*}{\textwidth}{@{\extracolsep{\fill}}ccccc}
\hline
\hline
\textbf{7837.81+i91.45}& $\Xi_{bc} \rho$ & $\Xi_{bc} \omega$ & $\Sigma_c \bar B^\ast$ & $\Sigma_b D^\ast$ \\
\hline
 $g_i$                 &0.04+i0.49&1.42+i0.49&0&-0.05+i0.01\\
 $g_iG^{\uppercase\expandafter{\romannumeral2}}_i$&-24.78-i9.65&-58.19+i55.70&0&1.86-i2.19\\
\hline
                      & $\Omega_{bc}K^\ast$ & $\Xi_{bc}\phi$ & $\Xi_c^\prime\bar B_s^\ast$ & $\Xi_b^\prime D_s^\ast$\\
\hline
 $g_i$                &3.45-i1.23&-2.07-i0.61&0&0.14-i0.57\\
 $g_iG^{\uppercase\expandafter{\romannumeral2}}_i$&-40.99-i8.16&14.22+i12.01&0&-1.13+i1.81\\
 \hline
\end{tabular*}
\end{table}
\subsection{Vector-baryon$(1/2^+)$, mixed antisymmetric sector}
In Table~\ref{tab:vbma} we show the coupled channels and the thresholds and in Table~\ref{VBMA} we show the $D_{ij}$ coefficients.
The states are shown in Table~\ref{vbmapole}.
The couplings of the states to the different channels are shown in Tables~\ref{vbmapole1}, \ref{vbmapole2}, \ref{vbmapole3} and \ref{vbmapole4}.
The $7599\,\rm MeV$ and $7826\,\rm MeV$ states are basically the same states as before
because the dominant channel and decay channel appear in both the $MS$ and $MA$ representations.
\begin{table}[H]
\renewcommand\arraystretch{1.0}
\centering
\caption{\vadjust{\vspace{-2pt}}Channels considered for sector $J^P=1/2^-,3/2^-$ (MA).}\label{tab:vbma}
\begin{tabular*}{1.00\textwidth}{@{\extracolsep{\fill}}c|ccccccccc}
\hline
\hline
\textbf{Channel} & $\Lambda_c\bar B^\ast$ & $\Lambda_b D^\ast$ & $\Xi_{bc} \rho$ &
$\Xi_{bc} \omega$ & $\Omega_{bc}K^\ast$ & $\Xi_c \bar B_s^\ast$ & $\Xi_b D_s^\ast$ & $\Xi_{bc} \phi$\\
\hline
\textbf{Threshold (MeV)} & 7611 & 7629 & 7694 & 7702 & 7882 & 7884 & 7905 & 7938 \\
\hline
\hline
\end{tabular*}
\end{table}
\begin{table}[H]
\renewcommand\arraystretch{1.00}
\begin{adjustwidth}{0\textwidth}{0\textwidth}
\caption{\vadjust{\vspace{-2pt}}$D_{ij}$ coefficients for sector $J^P=1/2^-,3/2^-$ (MA).}\label{VBMA}
\begin{tabular*}{1.0\textwidth}{@{\extracolsep{\fill}}c|cccccccc}
\hline
\hline
\textbf{$J^P=1/2^-,3/2^-$} & $\Lambda_c\bar B^\ast$ & $\Lambda_b D^\ast$ & $\Xi_{bc} \rho$ &
$\Xi_{bc} \omega$ & $\Omega_{bc}K^\ast$ & $\Xi_c \bar B_s^\ast$ & $\Xi_b D_s^\ast$ & $\Xi_{bc} \phi$\\
\hline
$\Lambda_c\bar B^\ast$ & $-1$ & $\sim$0 & 0 & 0 & 0 & $-1$ & 0 & 0 \\
\hline
$\Lambda_b D^\ast$ &  & $-1$ & $-\frac{\sqrt 3}{2}\lambda$ & $-\frac{1}{2}\lambda$ & 0 & 0 & $-1$ & 0 \\
\hline
$\Xi_{bc} \rho$ &  &  & $-2$ & 0 & $\sqrt\frac{3}{2}$ & 0 & 0 & 0 \\
\hline
$\Xi_{bc} \omega$ &  &  &  & 0 & $-\frac{1}{\sqrt 2}$ & 0 & 0 & 0 \\
\hline
$\Omega_{bc}K^\ast$ &  &  &  &  & $-1$ & 0 & $\frac{1}{\sqrt 2}\lambda$ & 1 \\
\hline
$\Xi_c \bar B_s^\ast$ &  &  &  &  &  & $-1$ & $\sim$0 & 0 \\
\hline
$\Xi_b D_s^\ast$ &  &  &  &  &  &  & $-1$ & $-\frac{1}{\sqrt 2}\lambda$ \\
\hline
$\Xi_{bc} \phi$ &  &  &  &  &  &  &  & 0 \\
\hline
\hline
\end{tabular*}
\end{adjustwidth}
\end{table}
\begin{table}[H]
\renewcommand\arraystretch{1.0}
\centering
\caption{\vadjust{\vspace{-2pt}}Poles for vector-baryon(1/2) (MA) states (all units are in MeV).}\label{vbmapole}
\begin{tabular*}{1.00\textwidth}{@{\extracolsep{\fill}}ccccccc}
\hline
 $q_{\rm max}$    &600            &650            &700            &750            &800 \\
\hline
              &7538.75        &7508.55        &7471.42        &7427.31        &7376.44 \\
\hline
              &7559.38        &7531.22        &7497.76        &7459.40        &7416.60 \\
\hline
              &7621.95        &7599.65        &7575.06        &7548.85        &7521.72 \\
\hline
              &7853.18+i80.53 &7826.83+i77.82 &7798.50+i72.97 &7769.94+i65.32 &7740.93+i54.42 \\
\hline
\hline
\end{tabular*}
\end{table}
\begin{table}[H]
\renewcommand\arraystretch{1.0}
\centering
\caption{\vadjust{\vspace{-2pt}}Coupling constants to various channels
and $g_iG^{\uppercase\expandafter{\romannumeral2}}_i$ in MeV with $q_{\rm max}=650\,\rm MeV$.}\label{vbmapole1}
\begin{tabular*}{\textwidth}{@{\extracolsep{\fill}}ccccc}
\hline
\hline
\textbf{7508.55}& $\Lambda_c\bar B^\ast$ & $\Lambda_b D^\ast$ & $\Xi_{bc} \rho$ & $\Xi_{bc} \omega$  \\
\hline
 $g_i$                 &7.49&0&0&0\\
 $g_iG^{\uppercase\expandafter{\romannumeral2}}_i$&-18.51&0&0&0\\
\hline
                      & $\Omega_{bc}K^\ast$ & $\Xi_c \bar B_s^\ast$ & $\Xi_b D_s^\ast$ & $\Xi_{bc} \phi$\\
\hline
 $g_i$                &0&7.22&0&0\\
 $g_iG^{\uppercase\expandafter{\romannumeral2}}_i$&0&-6.94&0&0\\
 \hline
\end{tabular*}
\end{table}
\begin{table}[H]
\renewcommand\arraystretch{1.0}
\centering
\caption{\vadjust{\vspace{-2pt}}Coupling constants to various channels
and $g_iG^{\uppercase\expandafter{\romannumeral2}}_i$ in MeV with $q_{\rm max}=650\,\rm MeV$.}\label{vbmapole2}
\begin{tabular*}{\textwidth}{@{\extracolsep{\fill}}ccccc}
\hline
\hline
\textbf{7531.22}& $\Lambda_c\bar B^\ast$ & $\Lambda_b D^\ast$ & $\Xi_{bc} \rho$ & $\Xi_{bc} \omega$  \\
\hline
 $g_i$                 &0&4.32&1.43&0.18\\
 $g_iG^{\uppercase\expandafter{\romannumeral2}}_i$&0&-28.22&-12.03&-1.44\\
\hline
                      & $\Omega_{bc}K^\ast$ & $\Xi_c \bar B_s^\ast$ & $\Xi_b D_s^\ast$ & $\Xi_{bc} \phi$\\
\hline
 $g_i$                &-0.71&0&3.97&0.23\\
 $g_iG^{\uppercase\expandafter{\romannumeral2}}_i$&3.38&0&-9.64&-0.90\\
 \hline
\end{tabular*}
\end{table}
\begin{table}[H]
\renewcommand\arraystretch{1.0}
\centering
\caption{\vadjust{\vspace{-2pt}}Coupling constants to various channels
and $g_iG^{\uppercase\expandafter{\romannumeral2}}_i$ in MeV with $q_{\rm max}=650\,\rm MeV$.}\label{vbmapole3}
\begin{tabular*}{\textwidth}{@{\extracolsep{\fill}}ccccc}
\hline
\hline
\textbf{7599.65}& $\Lambda_c\bar B^\ast$ & $\Lambda_b D^\ast$ & $\Xi_{bc} \rho$ & $\Xi_{bc} \omega$  \\
\hline
 $g_i$                 &0&-0.74&3.35&-0.42\\
 $g_iG^{II}_i$&0&8.59&-36.94&4.55\\
\hline
                      & $\Omega_{bc}K^\ast$ & $\Xi_c \bar B_s^\ast$ & $\Xi_b D_s^\ast$ & $\Xi_{bc} \phi$\\
\hline
 $g_i$                &-2.30&0&-1.07&0.44\\
 $g_iG^{II}_i$&12.79&0&3.05&-1.99\\
 \hline
\end{tabular*}
\end{table}
\begin{table}[H]
\renewcommand\arraystretch{1.0}
\centering
\caption{\vadjust{\vspace{-2pt}}Coupling constants to various channels
and $g_iG^{\uppercase\expandafter{\romannumeral2}}_i$ in MeV with $q_{\rm max}=650\,\rm MeV$.}\label{vbmapole4}
\begin{tabular*}{\textwidth}{@{\extracolsep{\fill}}ccccc}
\hline
\hline
\textbf{7826.83+i77.82}& $\Lambda_c\bar B^\ast$ & $\Lambda_b D^\ast$ & $\Xi_{bc} \rho$ & $\Xi_{bc} \omega$  \\
\hline
 $g_i$                 &0&-0.09+i0.03&0.11+i0.47&1.36+i0.40\\
 $g_iG^{\uppercase\expandafter{\romannumeral2}}_i$&0&-1.43-i8.14&-23.63-i5.98&-50.30+i49.92\\
\hline
                      & $\Omega_{bc}K^\ast$ & $\Xi_c \bar B_s^\ast$ & $\Xi_b D_s^\ast$ & $\Xi_{bc} \phi$\\
\hline
 $g_i$                &3.23-i1.05&0&-1.87+i0.13&-2.07-i0.51\\
 $g_iG^{\uppercase\expandafter{\romannumeral2}}_i$&-38.27-i5.14&0&10.68+i4.71&15.03+i9.96\\
 \hline
\end{tabular*}
\end{table}
\subsection{Pseudoscalar-baryon$(3/2)^+$ states}
In this case the spin wave function, $\chi_S$, is full symmetric and the states generated are in the sector $J^P=3/2^-$.
In Table~\ref{tab:pbstar} we show the channels and the thresholds and in Table~\ref{PBstar} the $D_{ij}$ coefficients.
We observe three states, one with a width of about $190\,\rm MeV$ and the other two narrow.
Inspecting Tables~\ref{pbstarpole1}, \ref{pbstarpole2}, \ref{pbstarpole3} we can see that the first state couples strongly to $\Xi_{bc}^\ast\pi$,
which is open, and this is the reason for the large width. The second state couples mostly to $\Sigma_b^\ast D$ and the third one to $\Sigma_c^\ast\bar B$.
\begin{table}[H]
\renewcommand\arraystretch{1.0}
\centering
\caption{\vadjust{\vspace{-2pt}}Channels considered for sector $J^P=3/2^-$ (S).}\label{tab:pbstar}
\begin{tabular*}{1.00\textwidth}{@{\extracolsep{\fill}}c|cccccccc}
\hline
\hline
\textbf{Channel} & $\Xi_{bc}^*\pi$ & $\Xi_{bc}^*\eta$ & $\Omega_{bc}^*K$ & $\Sigma_b^* D$ & $\Sigma_c^*\bar B$ & $\Xi_b^*D_s$ & $\Xi_c^*\bar B_s$  \\
\hline
\textbf{Threshold (MeV)} & 7124 & 7534 & 7542 & 7701 & 7797 & 7921 & 8013 & \\
\hline
\hline
\end{tabular*}
\end{table}
\begin{table}[H]
\renewcommand\arraystretch{1.00}
\centering
\caption{\vadjust{\vspace{-2pt}}$D_{ij}$ coefficients for sector $J^P=3/2^-$ (S).}\label{PBstar}
\begin{tabular*}{1.00\textwidth}{@{\extracolsep{\fill}}c|cccccccc}
\hline
\hline
\textbf{$J^P=3/2^-$} & $\Xi_{bc}^*\pi$ & $\Xi_{bc}^*\eta$ & $\Omega_{bc}^*K$ & $\Sigma_b^* D$ & $\Sigma_c^*\bar B$ & $\Xi_b^*D_s$ & $\Xi_c^*\bar B_s$  \\
\hline
$\Xi_{bc}^*\pi$ & $-2$ & 0 & $\sqrt\frac{3}{2}$ & $\frac{1}{2}\lambda$ & 0 & 0 & 0 \\
\hline
$\Xi_{bc}^*\eta$ &  & 0 & $-\frac{2}{\sqrt 3}$ & $-\frac{1}{\sqrt 2}\lambda$ & 0 & $-\frac{1}{\sqrt 6}\lambda$ & 0 \\
\hline
$\Omega_{bc}^*K$ &  &  & $-1$ & 0 & 0 & $\frac{1}{\sqrt 2}\lambda$ & 0\\
\hline
$\Sigma_b^* D$ &  &  &  & $-3$ & 0 & $\sqrt 3$ & 0 \\
\hline
$\Sigma_c^*\bar B$ &  &  &  &  & $-3$ & 0 & $\sqrt 3$ \\
\hline
$\Xi_b^*D_s$ &  &  &  &  &  & $-1$ & 0 \\
\hline
$\Xi_c^*\bar B_s$ &  &  &  &  &  &  & $-1$  \\
\hline
\hline
\end{tabular*}
\end{table}
\begin{table}[H]
\renewcommand\arraystretch{1.0}
\centering
\caption{\vadjust{\vspace{-2pt}}Poles for pesudoscalar-baryon(3/2) (S) states (all units are in MeV).}\label{pbstarpole}
\begin{tabular*}{1.00\textwidth}{@{\extracolsep{\fill}}ccccccc}
\hline
 $q_{\rm max}$    &600            &650            &700            &750            &800 \\
\hline
              &7198.92+i103.55&7198.33+i94.56 &7196.86+i85.55 &7195.05+i76.23 &7195.23+i68.21 \\
\hline
              &7455.11+i0.49  &7393.03+i0.76  &7326.17+i1.31  &7531.93+i7.50  &7180.43+i2.88 \\
\hline
              &7513.15        &7434.97        &7347.43        &7251.46        &7148.17 \\
\hline
\hline
\end{tabular*}
\end{table}
\begin{table}[H]
\renewcommand\arraystretch{1.0}
\centering
\caption{\vadjust{\vspace{-2pt}}Coupling constants to various channels
and $g_iG^{\uppercase\expandafter{\romannumeral2}}_i$ in MeV with $q_{\rm max}=650\,\rm MeV$.}\label{pbstarpole1}
\begin{tabular*}{\textwidth}{@{\extracolsep{\fill}}ccccc}
\hline
\hline
\textbf{7198.33+i94.56}& $\Xi_{bc}^*\pi$ & $\Xi_{bc}^*\eta$ & $\Omega_{bc}^*K$ & $\Sigma_b^* D$  \\
\hline
 $g_i$                 &1.70+i1.23&-0.03-i0.11&-0.87-i0.74&-0.74-i0.50\\
 $g_iG^{\uppercase\expandafter{\romannumeral2}}_i$&-73.77-i12.77&0.06+i0.68&4.63+i5.57&1.36+i1.27\\
\hline
                      & $\Sigma_c^*\bar B$ & $\Xi_b^*D_s$ & $\Xi_c^*\bar B_s$ &\\
\hline
 $g_i$                &0&0.21+i0.25&0&\\
 $g_iG^{\uppercase\expandafter{\romannumeral2}}_i$&0&-0.26-i0.40&0&\\
 \hline
\end{tabular*}
\end{table}
\begin{table}[H]
\renewcommand\arraystretch{1.0}
\centering
\caption{\vadjust{\vspace{-2pt}}Coupling constants to various channels
and $g_iG^{\uppercase\expandafter{\romannumeral2}}_i$ in MeV with $q_{\rm max}=650\,\rm MeV$.}\label{pbstarpole2}
\begin{tabular*}{\textwidth}{@{\extracolsep{\fill}}ccccc}
\hline
\hline
\textbf{7393.03+i0.76}& $\Xi_{bc}^*\pi$ & $\Xi_{bc}^*\eta$ & $\Omega_{bc}^*K$ & $\Sigma_b^* D$  \\
\hline
 $g_i$                 &-0.02+i0.16&0.25-i0.03&0.05-i0.10&9.41-i0.03\\
 $g_iG^{\uppercase\expandafter{\romannumeral2}}_i$&-4.32-i2.86&-2.60+i0.27&-0.54+i1.07&-29.44+i0.04\\
\hline
                      & $\Sigma_c^*\bar B$ & $\Xi_b^*D_s$ & $\Xi_c^*\bar B_s$ &\\
\hline
 $g_i$                &0&-5.21+i0.01&0&\\
 $g_iG^{\uppercase\expandafter{\romannumeral2}}_i$&0&10.10-i0.01&0&\\
 \hline
\end{tabular*}
\end{table}
\begin{table}[H]
\renewcommand\arraystretch{1.0}
\centering
\caption{\vadjust{\vspace{-2pt}}Coupling constants to various channels
and $g_iG^{\uppercase\expandafter{\romannumeral2}}_i$ in MeV with $q_{\rm max}=650\,\rm MeV$.}\label{pbstarpole3}
\begin{tabular*}{\textwidth}{@{\extracolsep{\fill}}ccccc}
\hline
\hline
\textbf{7434.97}& $\Xi_{bc}^*\pi$ & $\Xi_{bc}^*\eta$ & $\Omega_{bc}^*K$ & $\Sigma_b^* D$  \\
\hline
 $g_i$                 &0&0&0&0\\
 $g_iG^{\uppercase\expandafter{\romannumeral2}}_i$&0&0&0&0\\
\hline
                      & $\Sigma_c^*\bar B$ & $\Xi_b^*D_s$ & $\Xi_c^*\bar B_s$ &\\
\hline
 $g_i$                &18.14&0&-10.20&\\
 $g_iG^{\uppercase\expandafter{\romannumeral2}}_i$&-18.47&0&6.94&\\
 \hline
\end{tabular*}
\end{table}

\section{Wave functions}
It is interesting to see how are the wave functions that we have generated.
They are a bit different than ordinary wave functions with local potentials decreasing very rapidly as $r \to \infty$.
In order to see that, we have to go back to the work of Ref.~\cite{Gamermann:2009uq},
where we find that the use of Eq.~\eqref{eq:BS} with a $G$ function regularized with a cut off $q_{\rm max}$ is equivalent to
solving the Lippmann-Schwinger equation with a potential
\begin{equation}\label{eq:LSeq}
  V(\vec q, \vec q^{\,'}) = V \,\theta (q_{\rm max}-|\vec q\,|)\; \theta (q_{\rm max}-|\vec q^{\,'}|),
\end{equation}
which is a non-local potential.
The wave function in momentum space is particularly easy (see Eqs.~(34), (47) of Ref.~\cite{Gamermann:2009uq}
and Eq.~(105) of Ref.~\cite{yamagata} generalizing to relativistic energies)
\begin{equation}
  \langle \vec q\, | \psi \rangle =g \frac{\theta (q_{\rm max}-|\vec q\,|)}{E-w_1(\vec q\,)-w_2(\vec q\,)},
\end{equation}
where $w_i(\vec q\,)= \sqrt{m_i^2+ \vec q^{\, 2}}$ and $g$ is the coupling of the wave function to the channel with particles 1 and 2.

The wave function in coordinate space is given by
\begin{equation}\label{eq:wavFun}
\langle \vec x \,| \psi \rangle = \int \frac{d^3q}{(2\pi)^{3/2}} \, e^{i\vec q \cdot \vec x} \, \langle \vec q \,| \psi\rangle
=\frac{2\pi}{(2\pi)^{3/2}} \, g \, \frac{2}{r} \int_0^{q_{\rm max}} q\, dq \, \frac{1}{E-w_1(\vec q\,)-w_2(\vec q\,)} \, \sin (qr).
\end{equation}

It is interesting to note that unlike in ordinary potentials,
which could be simulated with $q_{\rm max} \to \infty$ and $w_1(\vec q\,)+w_2(\vec q\,)$ and $\sin (qr)$ providing convergence in the $q$ integration,
in our case $q_{\rm max}=650\, {\rm MeV}$ corresponds to a value where $\vec q^{\, 2} / m_i^2 $ is very small
and the $w_1(\vec q\,)+w_2(\vec q\,)$ term does not help in the convergence of the integral,
which is determined by $q_{\rm max}$.
One can then see that the shape of the wave function does not depend much on $E$,
which diverts from ordinary wave functions with rapidly decreasing local potential where the size is roughly given by $r= 1/ \sqrt{2\mu B}$,
with $\mu$ the reduced mass and $B$ the binding.
We show this in an example of a very bound component,
the $7372 \, {\rm MeV}$ state of Table \ref{pbmspole2},
which couples mostly to $\Sigma_b D$.

\begin{figure}[H]
  \centering
  \includegraphics[width=0.55\textwidth]{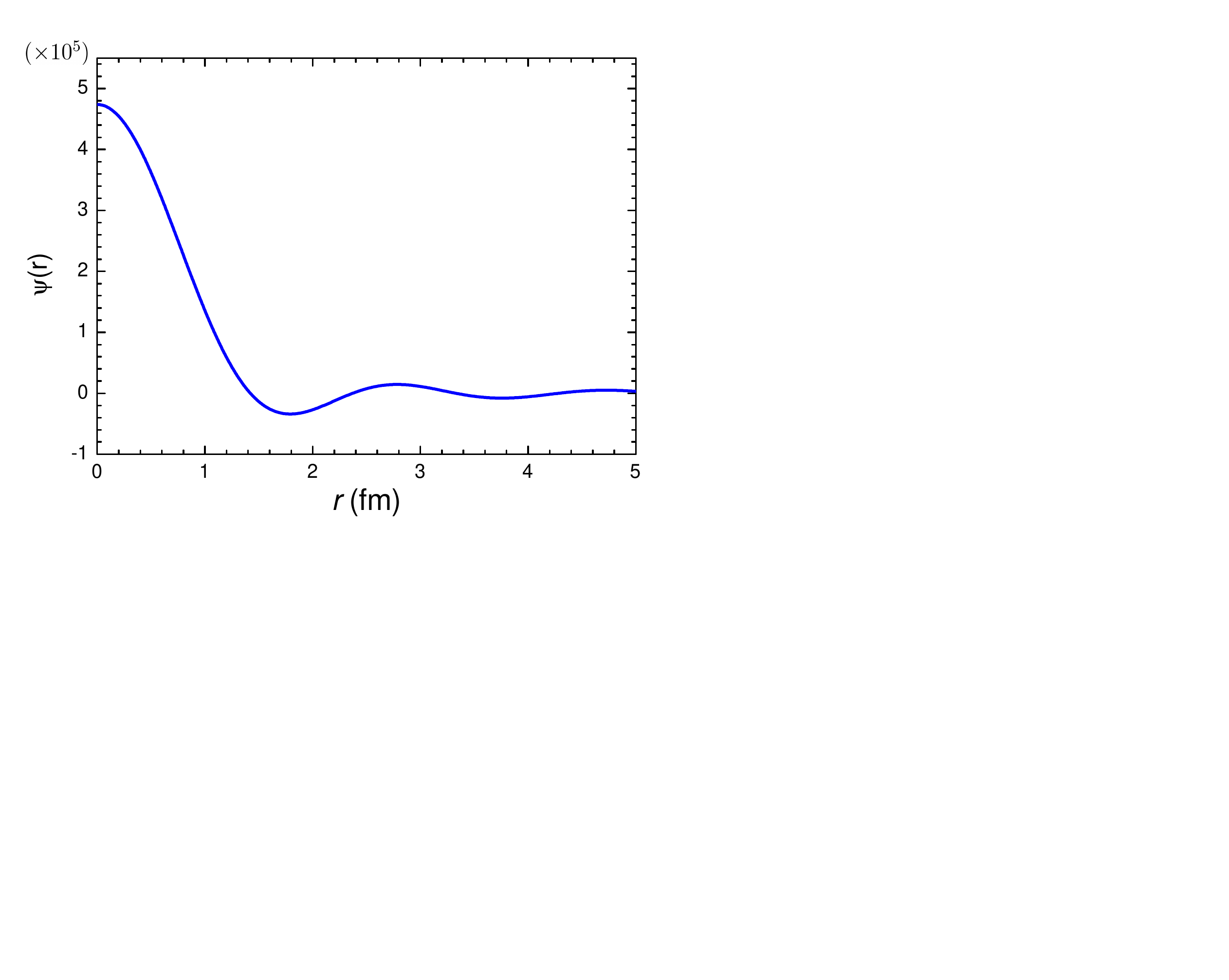}
  \caption{Wave function $\Psi (r) $ corresponding to channel $\Sigma_b D$.}
  \label{Fig:WaveFun}
\end{figure}

\begin{figure}[H]
  \centering
  \includegraphics[width=0.55\textwidth]{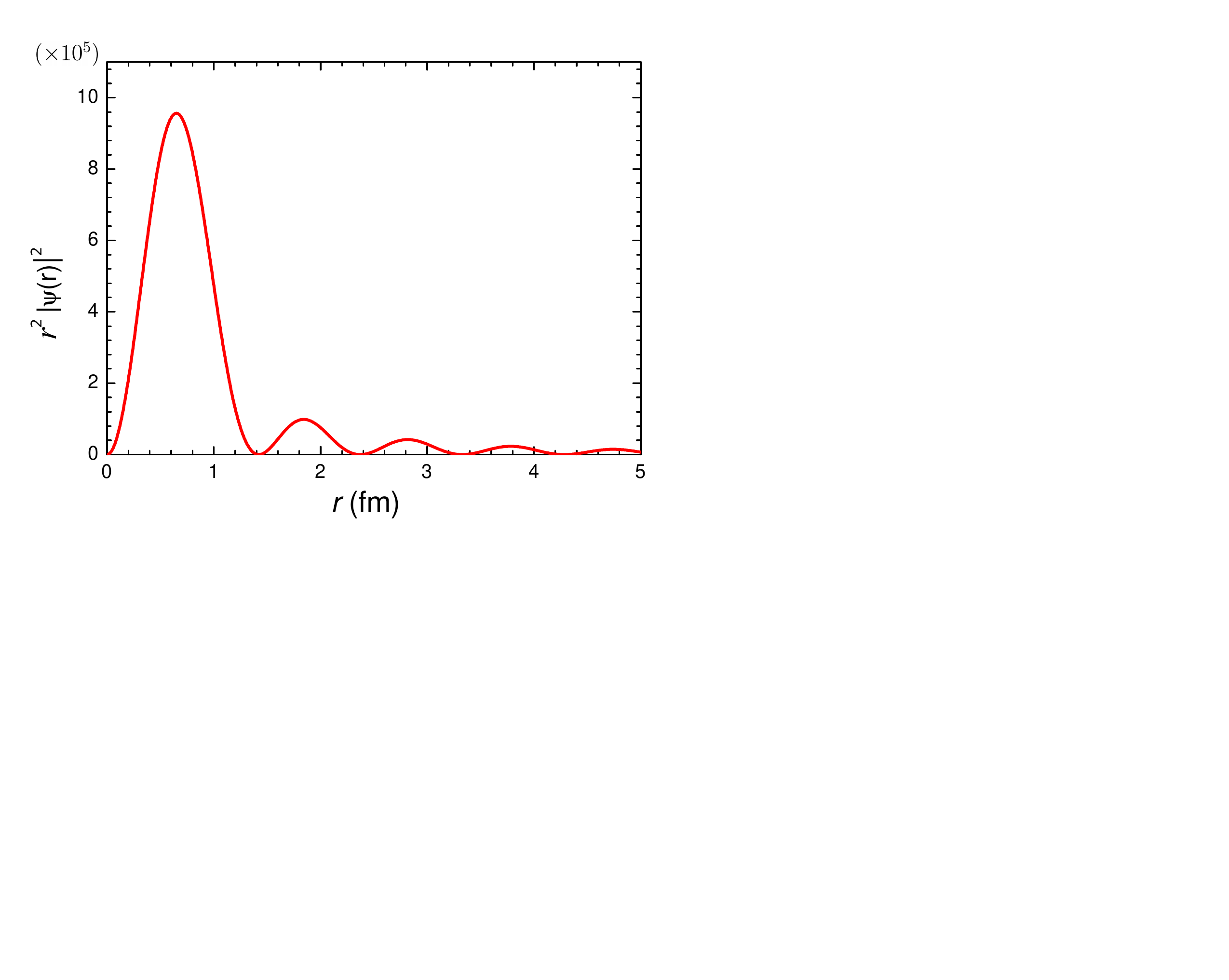}
  \caption{The function $r^2\, |\Psi (r)|^2 $ corresponding to channel $\Sigma_b D$.}
  \label{Fig:WaveFunSqr}
\end{figure}

In Fig.~\ref{Fig:WaveFun} we show the wave function corresponding to this channel.
While naively we would expect a size of around $0.2\,$ fm according to the intuitive formula,
we find that the wave function extends much further and even at $r\simeq 1\, {\rm fm}$ is not negligible.
This can be better appreciated in Fig.~\ref{Fig:WaveFunSqr} where we plot the wave function squared times $r^2$.
We see that it peaks around $0.7\, {\rm fm}$ and still has a sizable strength around $1\, {\rm fm}$ and beyond.

\section{Conclusions}
We have studied the meson-baryon interaction in the sector corresponding to $\Xi_{bc}$ quantum numbers.
We take the coupled channels that can lead to these quantum numbers and study their interaction in $s$-wave.
The model used for the interaction is based on an extrapolation of the local hidden gauge approach,
which uses vector exchange as the source of interaction.
The dominant terms come from the exchange of light vector mesons
which render the heavy quarks as spectators and the approach automatically satisfies the heavy quark symmetry rules.
The interaction is properly unitarized in coupled channels and by looking at poles in the second Riemann sheet we look for the states of the system.
We consider the interaction of pseudoscalars with baryons of $J^P=1/2^+,3/2^+$
and of vectors with baryons of $J^P=1/2^+$ and distinguish spin mixed symmetric and mixed antisymmetric states in analogy to the $\Xi$ and $\Xi^\prime$ states.
We find several states which correspond to bound meson-baryon states with zero or small widths and a few that have a large width,
Since the input used to generate these states is the same one used to study $\Omega_c$ and states of hidden charm,
which produced results in excellent agreement with experiment,
we have confidence that the predictions made are realistic and encourage the experimental search of such states.
In some cases the main decay channels have been identified and this can be useful in the planning of experimental proposals.

\section{ACKNOWLEDGEMENT}
We thank the hospitality of Guangxi Normal University where the main part of this work is done.
Qi-Xin Yu acknowledges the support from the National Natural Science Foundation of China (Grants No.~11775024, No.~11575023 and No.~11805153)
and China Scholarship Council.
This work is partly supported by the National Natural Science Foundation of China under Grants Nos.~11975083, 11847317 and 11565007.
This work is partly supported by the Spanish Ministerio de Economia y Competitividad and European FEDER funds under Contracts No.~FIS2017-84038-C2-1-P B and No.~FIS2017-84038-C2-2-P B, and the Generalitat Valenciana in the program Prometeo II-2014/068, and the project Severo Ochoa of IFIC, SEV-2014-0398.


\begin{thebibliography}{}
\bibitem{Aaij:2015tga}
  R.~Aaij {\it et al.} [LHCb Collaboration],
  Phys.\ Rev.\ Lett.\  {\bf 115}, 072001 (2015).

\bibitem{Aaij:2019vzc}
  R.~Aaij {\it et al.} [LHCb Collaboration],
  Phys.\ Rev.\ Lett.\  {\bf 122}, no. 22, 222001 (2019).

\bibitem{Aaij:2017nav}
  R.~Aaij {\it et al.} [LHCb Collaboration],
  Phys.\ Rev.\ Lett.\  {\bf 118}, no. 18, 182001 (2017).

\bibitem{Aaij:2017ueg}
  R.~Aaij {\it et al.} [LHCb Collaboration],
  Phys.\ Rev.\ Lett.\  {\bf 119}, no. 11, 112001 (2017).

\bibitem{DeRujula:1975qlm}
  A.~De Rujula, H.~Georgi and S.~L.~Glashow,
  Phys.\ Rev.\ D {\bf 12}, 147 (1975).

\bibitem{Ebert:2002ig}
  D.~Ebert, R.~N.~Faustov, V.~O.~Galkin and A.~P.~Martynenko,
  Phys.\ Rev.\ D {\bf 66}, 014008 (2002).

\bibitem{Korner:1994nh}
  J.~G.~K{\"o}rner, M.~Kramer and D.~Pirjol,
  Prog.\ Part.\ Nucl.\ Phys.\  {\bf 33}, 787 (1994).

\bibitem{Chen:2016spr}
  H.~X.~Chen, W.~Chen, X.~Liu, Y.~R.~Liu and S.~L.~Zhu,
  Rept.\ Prog.\ Phys.\  {\bf 80}, no. 7, 076201 (2017).

\bibitem{Yu:2017zst}
  F.~S.~Yu, H.~Y.~Jiang, R.~H.~Li, C.~D.~L{\"u}, W.~Wang and Z.~X.~Zhao,
  Chin.\ Phys.\ C {\bf 42}, no. 5, 051001 (2018).

\bibitem{Wang:2017mqp}
  W.~Wang, F.~S.~Yu and Z.~X.~Zhao,
  Eur.\ Phys.\ J.\ C {\bf 77}, no. 11, 781 (2017).

\bibitem{Wang:2017azm}
  W.~Wang, Z.~P.~Xing and J.~Xu,
  Eur.\ Phys.\ J.\ C {\bf 77}, no. 11, 800 (2017).

\bibitem{Gutsche:2017hux}
  T.~Gutsche, M.~A.~Ivanov, J.~G.~K{\"o}rner and V.~E.~Lyubovitskij,
  Phys.\ Rev.\ D {\bf 96}, no. 5, 054013 (2017).

\bibitem{Sharma:2017txj}
  N.~Sharma and R.~Dhir,
  Phys.\ Rev.\ D {\bf 96}, no. 11, 113006 (2017).

\bibitem{Hu:2017dzi}
  X.~H.~Hu, Y.~L.~Shen, W.~Wang and Z.~X.~Zhao,
  Chin.\ Phys.\ C {\bf 42}, no. 12, 123102 (2018).

\bibitem{Shi:2017dto}
  Y.~J.~Shi, W.~Wang, Y.~Xing and J.~Xu,
  Eur.\ Phys.\ J.\ C {\bf 78}, no. 1, 56 (2018).

\bibitem{Karliner:2018hos}
  M.~Karliner and J.~L.~Rosner,
  Phys.\ Rev.\ D {\bf 97}, no. 9, 094006 (2018).

\bibitem{Zhao:2018mrg}
  Z.~X.~Zhao,
  Eur.\ Phys.\ J.\ C {\bf 78}, no. 9, 756 (2018).

\bibitem{Xing:2018lre}
  Z.~P.~Xing and Z.~X.~Zhao,
  Phys.\ Rev.\ D {\bf 98}, no. 5, 056002 (2018).

\bibitem{Wang:2018lhz}
  Z.~G.~Wang,
  Eur.\ Phys.\ J.\ C {\bf 78}, no. 10, 826 (2018).

\bibitem{Cheng:2018mwu}
  H.~Y.~Cheng and Y.~L.~Shi,
  Phys.\ Rev.\ D {\bf 98}, no. 11, 113005 (2018).

\bibitem{Jiang:2018oak}
  L.~J.~Jiang, B.~He and R.~H.~Li,
  Eur.\ Phys.\ J.\ C {\bf 78}, no. 11, 961 (2018).

\bibitem{Zhang:2018llc}
  Q.~A.~Zhang,
  Eur.\ Phys.\ J.\ C {\bf 78}, no. 12, 1024 (2018).

\bibitem{Gutsche:2018msz}
  T.~Gutsche, M.~A.~Ivanov, J.~G.~K{\"o}rner, V.~E.~Lyubovitskij and Z.~Tyulemissov,
  Phys.\ Rev.\ D {\bf 99}, no. 5, 056013 (2019).

\bibitem{Ridgway:2019zks}
  A.~K.~Ridgway and M.~B.~Wise,
  Phys.\ Lett.\ B {\bf 793}, 181 (2019).

\bibitem{Cheng:2019sxr}
  H.~Y.~Cheng and F.~Xu,
  Phys.\ Rev.\ D {\bf 99}, no. 7, 073006 (2019).

\bibitem{Gerasimov:2019jwp}
  A.~S.~Gerasimov and A.~V.~Luchinsky,
  arXiv:1905.11740 [hep-ph].

\bibitem{Li:2017pxa}
  H.~S.~Li, L.~Meng, Z.~W.~Liu and S.~L.~Zhu,
  Phys.\ Lett.\ B {\bf 777}, 169 (2018).

\bibitem{Xiao:2017udy}
  L.~Y.~Xiao, K.~L.~Wang, Q.~f.~Lu, X.~H.~Zhong and S.~L.~Zhu,
  Phys.\ Rev.\ D {\bf 96}, no. 9, 094005 (2017).

\bibitem{Mehen:2017nrh}
  T.~Mehen,
  Phys.\ Rev.\ D {\bf 96}, no. 9, 094028 (2017).

\bibitem{Cui:2017udv}
  E.~L.~Cui, H.~X.~Chen, W.~Chen, X.~Liu and S.~L.~Zhu,
  Phys.\ Rev.\ D {\bf 97}, no. 3, 034018 (2018).

\bibitem{Xiao:2017dly}
  L.~Y.~Xiao, Q.~F.~L{\"u} and S.~L.~Zhu,
  Phys.\ Rev.\ D {\bf 97}, no. 7, 074005 (2018).

\bibitem{Bahtiyar:2018vub}
  H.~Bahtiyar, K.~U.~Can, G.~Erkol, M.~Oka and T.~T.~Takahashi,
  Phys.\ Rev.\ D {\bf 98}, no. 11, 114505 (2018).

\bibitem{Li:2017cfz}
  H.~S.~Li, L.~Meng, Z.~W.~Liu and S.~L.~Zhu,
  Phys.\ Rev.\ D {\bf 96}, no. 7, 076011 (2017).

\bibitem{Meng:2017dni}
  L.~Meng, H.~S.~Li, Z.~W.~Liu and S.~L.~Zhu,
  Eur.\ Phys.\ J.\ C {\bf 77}, no. 12, 869 (2017).

\bibitem{Ozdem:2018uue}
  U.~{\"O}zdem,
  J.\ Phys.\ G {\bf 46}, no. 3, 035003 (2019).

\bibitem{Ozdem:2019zis}
  Ula\c{s} {\"O}zdem,
  arXiv:1906.08353 [hep-ph].

\bibitem{Wang:2017qvg}
  C.~Y.~Wang, C.~Meng, Y.~Q.~Ma and K.~T.~Chao,
  Phys.\ Rev.\ D {\bf 99}, no. 1, 014018 (2019).

\bibitem{Wang:2018ihk}
  Z.~G.~Wang,
  Eur.\ Phys.\ J.\ C {\bf 78}, no. 4, 300 (2018).

\bibitem{Azizi:2018dva}
  K.~Azizi, Y.~Sarac and H.~Sundu,
  Phys.\ Rev.\ D {\bf 98}, no. 5, 054002 (2018).

\bibitem{Aliev:2019lvd}
  T.~M.~Aliev and S.~Bilmis,
  Nucl.\ Phys.\ A {\bf 984}, 99 (2019).

\bibitem{Shi:2019hbf}
  Y.~J.~Shi, W.~Wang and Z.~X.~Zhao,
  arXiv:1902.01092 [hep-ph].

\bibitem{Shi:2019fph}
  Y.~J.~Shi, Y.~Xing and Z.~X.~Zhao,
  Eur.\ Phys.\ J.\ C {\bf 79}, no. 6, 501 (2019).

\bibitem{Mathur:2018rwu}
  N.~Mathur and M.~Padmanath,
  Phys.\ Rev.\ D {\bf 99}, no. 3, 031501 (2019).

\bibitem{Kiselev:2017eic}
  V.~V.~Kiselev, A.~V.~Berezhnoy and A.~K.~Likhoded,
  Phys.\ Atom.\ Nucl.\  {\bf 81}, no. 3, 369 (2018)
  [Yad.\ Fiz.\  {\bf 81}, no. 3, 356 (2018)].

\bibitem{Lu:2017meb}
  Q.~F.~L{\"u}, K.~L.~Wang, L.~Y.~Xiao and X.~H.~Zhong,
  Phys.\ Rev.\ D {\bf 96}, no. 11, 114006 (2017).

\bibitem{Shah:2017jkr}
  Z.~Shah and A.~K.~Rai,
  Eur.\ Phys.\ J.\ A {\bf 53}, no. 10, 195 (2017).

\bibitem{Zhou:2018pcv}
  Q.~S.~Zhou, K.~Chen, X.~Liu, Y.~R.~Liu and S.~L.~Zhu,
  Phys.\ Rev.\ C {\bf 98}, no. 4, 045204 (2018).

\bibitem{Weng:2018mmf}
  X.~Z.~Weng, X.~L.~Chen and W.~Z.~Deng,
  Phys.\ Rev.\ D {\bf 97}, no. 5, 054008 (2018).

\bibitem{Richard:2018yrm}
  J.~M.~Richard, A.~Valcarce and J.~Vijande,
  Phys.\ Rev.\ C {\bf 97}, no. 3, 035211 (2018).

\bibitem{Li:2019ekr}
  Q.~Li, C.~H.~Chang, S.~X.~Qin and G.~L.~Wang,
  arXiv:1903.02282 [hep-ph].

\bibitem{Albertus:2006ya}
  C.~Albertus, E.~Hernandez, J.~Nieves and J.~M.~Verde-Velasco,
  Eur.\ Phys.\ J.\ A {\bf 32}, 183 (2007),
  Erratum: [Eur.\ Phys.\ J.\ A {\bf 36}, 119 (2008)].

\bibitem{Shah:2017liu}
  Z.~Shah and A.~K.~Rai,
  Eur.\ Phys.\ J.\ C {\bf 77}, no. 2, 129 (2017).

\bibitem{Ma:2017nik}
  Y.~L.~Ma and M.~Harada,
  J.\ Phys.\ G {\bf 45}, no. 7, 075006 (2018).

\bibitem{Liu:2018bkx}
  M.~Z.~Liu, T.~W.~Wu, J.~J.~Xie, M.~Pavon Valderrama and L.~S.~Geng,
  Phys.\ Rev.\ D {\bf 98}, no. 1, 014014 (2018).

\bibitem{Mehen:2019cxn}
  T.~C.~Mehen and A.~Mohapatra,
  Phys.\ Rev.\ D {\bf 100}, no. 7, 076014 (2019).

\bibitem{Guo:2017vcf}
  Z.~H.~Guo,
  Phys.\ Rev.\ D {\bf 96}, no. 7, 074004 (2017).

\bibitem{Yan:2018zdt}
  M.~J.~Yan, X.~H.~Liu, S.~Gonz{\'a}lez-Sol{\'i}s, F.~K.~Guo, C.~Hanhart, U.~G.~Mei{\ss}ner and B.~S.~Zou,
  Phys.\ Rev.\ D {\bf 98}, no. 9, 091502 (2018).

\bibitem{Meng:2018zbl}
  L.~Meng and S.~L.~Zhu,
  Phys.\ Rev.\ D {\bf 100}, no. 1, 014006 (2019).

\bibitem{Chen:2017jjn}
  R.~Chen, A.~Hosaka and X.~Liu,
  Phys.\ Rev.\ D {\bf 96}, no. 11, 114030 (2017).

\bibitem{Yao:2018ifh}
  D.~L.~Yao,
  Phys.\ Rev.\ D {\bf 97}, no. 3, 034012 (2018).

\bibitem{Blin:2018pmj}
  A.~N.~Hiller Blin, Z.~F.~Sun and M.~J.~Vicente Vacas,
  Phys.\ Rev.\ D {\bf 98}, no. 5, 054025 (2018).

\bibitem{qixin}
  Q.~X.~Yu and X.~H.~Guo,
  Nucl.\ Phys.\ B {\bf 947}, 114727 (2019).

\bibitem{Olsen:2017bmm}
  S.~L.~Olsen, T.~Skwarnicki and D.~Zieminska,
  Rev.\ Mod.\ Phys.\  {\bf 90}, no. 1, 015003 (2018).

\bibitem{Karliner:2017qhf}
  M.~Karliner, J.~L.~Rosner and T.~Skwarnicki,
  Ann.\ Rev.\ Nucl.\ Part.\ Sci.\  {\bf 68}, 17 (2018).

\bibitem{Liu:2019zoy}
  Y.~R.~Liu, H.~X.~Chen, W.~Chen, X.~Liu and S.~L.~Zhu,
  Prog.\ Part.\ Nucl.\ Phys.\  {\bf 107}, 237 (2019).

\bibitem{Dias:2018qhp}
  J.~M.~Dias, V.~R.~Debastiani, J.-J.~Xie and E.~Oset,
  Phys.\ Rev.\ D {\bf 98}, no. 9, 094017 (2018).

\bibitem{Debastiani:2017ewu}
  V.~R.~Debastiani, J.~M.~Dias, W.~H.~Liang and E.~Oset,
  Phys.\ Rev.\ D {\bf 97}, no. 9, 094035 (2018).

\bibitem{Sakai:2017avl}
  S.~Sakai, L.~Roca and E.~Oset,
  Phys.\ Rev.\ D {\bf 96}, no. 5, 054023 (2017).

\bibitem{Bando:1984ej}
  M.~Bando, T.~Kugo, S.~Uehara, K.~Yamawaki and T.~Yanagida,
  Phys.\ Rev.\ Lett.\  {\bf 54}, 1215 (1985).

\bibitem{Bando:1987br}
  M.~Bando, T.~Kugo and K.~Yamawaki,
  Phys.\ Rept.\  {\bf 164}, 217 (1988).

\bibitem{Meissner:1987ge}
  U.~G.~Mei{\ss}ner,
  Phys.\ Rept.\  {\bf 161}, 213 (1988).

\bibitem{Nagahiro:2008cv}
  H.~Nagahiro, L.~Roca, A.~Hosaka and E.~Oset,
  Phys.\ Rev.\ D {\bf 79}, 014015 (2009).

\bibitem{Ecker:1989yg}
  G.~Ecker, J.~Gasser, H.~Leutwyler, A.~Pich and E.~de Rafael,
  Phys.\ Lett.\ B {\bf 223}, 425 (1989).

\bibitem{Montana:2017kjw}
  G.~Monta{\~n}a, A.~Feijoo and {\`A}.~Ramos,
  Eur.\ Phys.\ J.\ A {\bf 54}, no. 4, 64 (2018).

\bibitem{Xiao:2019aya}
  C.~W.~Xiao, J.~Nieves and E.~Oset,
  Phys.\ Rev.\ D {\bf 100}, no. 1, 014021 (2019).

\bibitem{Liu:2019tjn}
  M.~Z.~Liu, Y.~W.~Pan, F.~Z.~Peng, M.~S{\'a}nchez S{\'a}nchez, L.~S.~Geng, A.~Hosaka and M.~Pavon Valderrama,
  Phys.\ Rev.\ Lett.\  {\bf 122}, no. 24, 242001 (2019).

\bibitem{Yu:2018yxl}
  Q.~X.~Yu, R.~Pavao, V.~R.~Debastiani and E.~Oset,
  Eur.\ Phys.\ J.\ C {\bf 79}, no. 2, 167 (2019).

\bibitem{Liang:2017ejq}
  W.~H.~Liang, J.~M.~Dias, V.~R.~Debastiani and E.~Oset,
  Nucl.\ Phys.\ B {\bf 930}, 524 (2018).


\bibitem{Close:1979bt}
  F.~E.~Close,
  ``An Introduction to Quarks and Partons,''
  (Academic Press, Cambrige, 1979).

\bibitem{slzhu}
  Z.~F.~Sun, J.~He, X.~Liu, Z.~G.~Luo and S.~L.~Zhu,
  Phys.\ Rev.\ D {\bf 84}, 054002 (2011).

\bibitem{uchino}
  W.~H.~Liang, T.~Uchino, C.~W.~Xiao and E.~Oset,
  Eur.\ Phys.\ J.\ A {\bf 51}, no. 2, 16 (2015).

\bibitem{uchidos}
  T.~Uchino, W.~H.~Liang and E.~Oset,
  Eur.\ Phys.\ J.\ A {\bf 52}, no. 3, 43 (2016).

\bibitem{hepen}
  J.~He,
  Eur.\ Phys.\ J.\ C {\bf 79}, no. 5, 393 (2019).

\bibitem{xliupen}
  R.~Chen, Z.~F.~Sun, X.~Liu and S.~L.~Zhu,
  Phys.\ Rev.\ D {\bf 100}, no. 1, 011502 (2019).

\bibitem{Pavon}
  M.~Pavon Valderrama,
  arXiv:1907.05294 [hep-ph].

\bibitem{genpen}
  M.~Z.~Liu, T.~W.~Wu, M.~S¨¢nchez S¨¢nchez, M.~P.~Valderrama, L.~S.~Geng and J.~J.~Xie,
  arXiv:1907.06093 [hep-ph].

\bibitem{npa}
  J.~A.~Oller and E.~Oset,
  Nucl.\ Phys.\ A {\bf 620}, 438 (1997);
  Erratum: [Nucl.\ Phys.\ A {\bf 652}, 407 (1999)].

\bibitem{sigma}
  J.~R.~Pelaez,
  Phys.\ Rept.\  {\bf 658}, 1 (2016).

\bibitem{mizobe}
  E.~Oset, H.~Toki, M.~Mizobe and T.~T.~Takahashi,
  Prog.\ Theor.\ Phys.\  {\bf 103}, 351 (2000).

\bibitem{Acetione}
  F.~Aceti, M.~Bayar, E.~Oset, A.~Martinez Torres, K.~P.~Khemchandani, J.~M.~Dias, F.~S.~Navarra and M.~Nielsen,
  Phys.\ Rev.\ D {\bf 90}, no. 1, 016003 (2014).

\bibitem{Acetidos}
  F.~Aceti, M.~Bayar, J.~M.~Dias and E.~Oset,
  Eur.\ Phys.\ J.\ A {\bf 50}, 103 (2014).

\bibitem{albersig}
  A.~Martinez Torres, K.~P.~Khemchandani and E.~Oset,
  Eur.\ Phys.\ J.\ A {\bf 36}, 211 (2008).




\bibitem{Mizutani:2006vq}
  T.~Mizutani and A.~Ramos,
  Phys.\ Rev.\ C {\bf 74}, 065201 (2006).

\bibitem{Gamermann:2009uq}
  D.~Gamermann, J.~Nieves, E.~Oset and E.~Ruiz Arriola,
  Phys.\ Rev.\ D {\bf 81}, 014029 (2010).

\bibitem{yamagata}
  J.~Yamagata-Sekihara, J.~Nieves and E.~Oset,
  Phys.\ Rev.\ D {\bf 83}, 014003 (2011).

\end{thebibliography}
\end{document}